\documentclass[11pt]{article}
\usepackage[centertags]{amsmath}
\usepackage{amsfonts} \usepackage{amssymb} \usepackage{amsthm}

\def\one{{\rm 1\kern -.9mm l}}                             %

\def\beq{\begin{equation}}
\def\eeq{\end{equation}}
\def\beqa{\begin{eqnarray}}
\def\eeqa{\end{eqnarray}}
\begin{document}
\begin{titlepage}
\rightline{DSF-18/2004} \rightline{NORDITA-2004-56
}\rightline{----------} \vskip 3.0cm \centerline{{\LARGE \bf
Brane-Inspired Orientifold Field Theories}} \vskip 1.0cm
\centerline{\bf Paolo Di Vecchia$^a$, Antonella Liccardo$^b$, Raffaele Marotta$^b$
and Franco Pezzella$^b$ } \vskip .8cm \centerline{\sl $^a$ NORDITA,
Blegdamsvej 17, DK-2100 Copenhagen \O, Denmark }
\centerline{e--mail: {\tt divecchi@alf.nbi.dk}}  \vskip .4cm
\centerline{\sl $^b$ Dipartimento di Scienze Fisiche,
Universit\`{a} di Napoli and INFN, Sezione di Napoli}
\centerline{\sl Via Cintia - Complesso Universitario M. Sant'
Angelo I-80126 Napoli, Italy} \centerline{e--mail: {\tt
name.surname@na.infn.it}}\vskip .4cm

\vskip 2cm
\begin{abstract}
In this paper we consider the gauge theory living on 
the world-volume of a stack
of $N$ D3-branes of Type 0B/$\Omega' I_{6}(-1)^{F_{L}}$  and of its
orbifolds $C^2/ \mathbb{Z}_2$ and $C^3/(\mathbb{Z}_2\times
\mathbb{Z}_2)$. The gauge theories obtained in the three cases are a brane
realization of ``orientifold field theories'' having the bosonic
sector common with ${\cal N}=4,2,1$ super Yang-Mills respectively. In these
non-supersymmetric theories, we investigate the possibility of
keeping the gauge/gravity correspondence that has revealed itself 
so successful
in the case of supersymmetric theories. In the open
string framework we compute the coefficient of the gauge kinetic
term showing that the perturbative behaviour of the orientifold
field theory can be obtained from the closed string channel in the
large $N$ limit, where the theory exhibits Bose-Fermi degeneracy.
\end{abstract}
\vfill  {\small Work partially supported by the European
Commission RTN Programme HPRN-CT-2000-00131 and by MIUR.}
\end{titlepage}

\newpage

\tableofcontents       %
\vskip 1cm             %
\section{Introduction}
\label{intro}

In the last years it has been shown in several cases that the
pertubative properties such as the chiral and scale
anomalies~\cite{d3}~\cite{KOW}~\cite{ANOMA} of less supersymmetric and
non-conformal theories living on D-branes can be obtained from
their correponding supergravity solutions~\footnote{For general
reviews on various approaches see Refs.~[4] $\div$ [8]. See also Ref.
\cite{OHTA}}. This came as a surprise because the supergravity
``dual'' solution was supposed to give a correct description of
the gauge theory for large values  of the 't Hooft
gauge coupling constant~\cite{MALDA}. The explanation of this fact
was given in Ref.~\cite{LMP} where it was shown, in the case of
the two orbifolds $C^2/\mathbb{Z}_2$ and $C^3/(\mathbb{Z}_2\times
\mathbb{Z}_2)$, that the contribution of the massless open string
states to the coefficient of the gauge kinetic term obtained from
the annulus diagram is exactly equal, under open/closed string
duality, to the contribution of the massless closed string states.
Actually it can also be shown that the contribution of the massive
states is identically zero giving no threshold
corrections~\footnote{We thank Jose' F. Morales for pointing this
out to
  us.}.

The previous results have been shown to be valid in the case of
theories that, although non-conformal,  preserve some
supersymmetry. Can they be extended to non-supersymmetric theories?
Obvious candidates for non-supersymmetric theories are the Type 0
ones that have been studied by constructing their supergravity
duals in Refs.~\cite{KT9811}, \cite{KT9812} and \cite{KT9901}. The
problem is, however, that they have a tachyon in the closed string
sector. Tachyon free orientifolds of these theories, called $0'$
theories, were introduced in Ref.~\cite{SAGNOTTI} and their
properties were extensively studied from different points of view
in Refs.~\cite{BFL9904} \cite{AA9912} \cite{DM0010}.  On the
other hand it is clear that the most promising theories to study
with these methods are the ones that are as close as possible to
supersymmetric theories.  Recently non-supersymmetric theories
have been studied that in the large number of colours are
equivalent to supersymmetric theories~\footnote{See the
  recent review by Armoni, Shifman and Veneziano~\cite{ASV} and References
  therein. In Ref.~\cite{SS0309} $1/N$ corrections are analysed.}.
In Refs. \cite{SAGNOTTI}, \cite{AA9912} and \cite{KSK},
non-supersymmetric gauge theories that are conformal in the planar
limit have been analyzed. In particular a large $N$ conformal
non-supersymmetric gauge theory may be obtained as the
world-volume theory of $N$ D$3$-branes of the orientifold $\Omega'
I_6 (-1)^{F_L}$ of 0B theory, where $\Omega'$ is the world-sheet
parity, $I_6$ the inversion of the coordinates orthogonal to the
world-volume of the D3-branes and $F_L$ is the space-time fermion
number
operator in the left sector. The gauge theory so constructed is an
example of ``orientifold field theory'' which in the large $N$ limit
is equivalent to ${\cal{N}}=4$ super Yang-Mills.

More recently some attention has been paid to the orientifold
field theories that contain a gluon and a fermion transforming
according  to the two-index symmetric or antisymmetric
representation of the gauge group $SU(N)$ ~\cite{ASV2} and that in
the large $N$ limit are equivalent to ${\cal{N}}=1$ SYM.

In this paper we review the brane construction of the ${\cal{N}}=4$
orientifold field theory, which is planar equivalent
(i.e. equivalent in the limit
$N\rightarrow\infty$ with $\lambda=g^2_{YM}N$ fixed) to ${\cal {N}}=4$
super Yang-Mills. By means of the orbifold projections
$C^2/ \mathbb{Z}_2$ and $C^3/(\mathbb{Z}_2\times \mathbb{Z}_2)$,
we give a complete string description of the orientifold field
theory whose spectrum has, in the large $N$ limit, the same number
of degrees of freedom as ${\cal{N}}=2,1$ super Yang-Mills.
The latter theory has been shown to be planar equivalent, both at perturbative and not-perturbative level, to ${\cal N}=1$ SYM \cite{ASV}. 

In the open string framework we compute the running coupling constant
and show that in the large $N$ limit, where the
Bose-Fermi degeneracy of the gauge theory is recovered, we can
obtain the perturbative behaviour of the orientifold field
theories also from the closed string channel. We see, however, that
the next to leading term in the large $N$ expansion of the $\beta$-function cannot be obtained
from the closed string channel. This means that gauge/gravity
correspondence holds only in the planar limit as far as the running
coupling constant is concerned. When we then consider the
$\theta$-angle we see that both the leading and the next to leading
terms can be equivalently determined from
the open and the closed string channel.
This follows from the fact that in the string framework  the
$\theta$-angle does not admit threshold corrections and thus it is
invariant under open/closed string duality.

The paper is organized as follows. In Sect. (\ref{0B}) we
summarize the main properties of Type 0 theories. Sect.
(\ref{orienti}) is devoted to the construction of the orientifold
$0B/(\Omega ' I_6 (-1)^{F_L})$. In the first subsection we
construct its open and closed string spectrum, in the second one
we compute the one-loop open string diagrams and finally in the
third one  we introduce an external field and we compute the
running coupling constant in both the open and closed string
channels finding agreement between them only for large values
of $N$, being  $U(N)$ the gauge group. In Sect.
(\ref{orbiorienti}) we consider the orbifolds $C^2/ \mathbb{Z}_2$
and $C^3/(\mathbb{Z}_2 {\times} \mathbb{Z}_2 )$ of the previous
orientifold obtaining non-supersymmetric gauge theories that in
the large $N$ limit reduce respectively to ${\cal{N}}=2$ and
${\cal{N}}=1$ SYM. Finally in the Appendix we perform
some calculations useful in Sect. (\ref{orienti}).


\section{Type 0B String Theory}
\label{0B}

In this section we summarize the properties of Type 0 string
theories. In particular we will discuss their spectrum and D
branes and compute the annulus diagram describing the interaction
between D branes.

Type 0 string theories are non supersymmetric string models
obtained by applying on the Neveu-Scharz-Ramond model for closed
strings  the following non-chiral diagonal projections:
\begin{eqnarray}
P_{\rm NS-NS}=\frac{1+(-1)^{F+\tilde{F} +G+\tilde{G}}}{2}\qquad
P_{\rm R-R}=\frac{1\pm(-1)^{F+\tilde{F} +G+\tilde{G}}}{2}
\label{gso}
\end{eqnarray}
where the upper [lower] sign corresponds to 0B [0A]. $F$ is
the world-sheet fermion number defined as
\begin{eqnarray}
F= \sum_{t=1/2}^\infty \psi_{-t} {\cdot} \psi_t -1
\label{fns78}
\end{eqnarray}
in the NS-NS sector and
\begin{eqnarray}
(-1)^{F} = \psi_{11}(-1)^{{F}_R}~~,~~
\psi_{11} \equiv 2^5 \psi_{0}^{0}
\psi_{0}^{1} \dots \psi_{0}^{9}~~,~~
F_{R} = \sum_{n=1}^{\infty} \psi_{-n} {\cdot}
\psi_{n} \label{fr79}
\end{eqnarray}
in the R-R sector, with analogous definitions for $\tilde{F}$ and $(-1)^{\tilde{F}}$.
$G$ is the superghost number operator defined as follows:
\begin{eqnarray}
G & = &  - \sum_{m=1/2}^{\infty} \left( \gamma_{-m} \beta_{m}+ \beta_{-m} \gamma_{m}
\right) \qquad \hspace{.5cm} \mbox{NS sector} \\
G & = &   -\gamma_{0} \beta_{0} - \sum_{m=1}^{\infty} \left( \gamma_{-m} \beta_{m} + \beta_{-m} \gamma_{m} \right) \qquad \mbox{R sector} 
\end{eqnarray}
In addition it is imposed that the fermionic NS-R and R-NS sectors are
eliminated from the physical spectrum, obtaining purely bosonic string
models. Their spectrum can be determined by keeping only the string
states that are
left invariant by the action of the operators given in
Eq. (\ref{gso}), i.e.:
\begin{eqnarray}
&&\mbox{Type 0A}\qquad ( {\rm NS}\,- \,,\,{\rm NS}\,-) \otimes ( {\rm
NS}\,+\,,\,{\rm NS}\,+)\otimes ( {\rm R}\,-\,,\,{\rm R}\,+)\otimes( {\rm R}\,+\,,\,{\rm R}\,-)\label{70as}\\
&& \mbox{Type 0B}\qquad ( {\rm NS}\,-\,,\,{\rm NS}\,-) \otimes ( {\rm
NS}\,+\,,\,{\rm NS}\,+)\otimes ( {\rm R}\,-\,,\,{\rm R}\,-)\otimes( {\rm R}\,+\,,\,{\rm R}\,+)\label{70bs}
\end{eqnarray}
where the signs in the various sectors refer to the values respectively taken
by $(-1)^{F}$ and $(-1)^{\bar{F}}$. In the (NS $-$ , NS $-$) sector the
lowest state is a tachyon, while the massless states live in the
(NS + , NS +) sector. In the picture $(-1,-1)$ they are described by:
\begin{equation}
\psi^{\mu}_{-\frac{1}{2}} {\tilde{\psi}}^{\nu}_{-\frac{1}{2}}
\hspace{.2cm} | 0,\, \tilde{0},\,k \rangle_{(-1,-1)}
\label{nsmsl9}
\end{equation}
and  are the  same as in Type II
theories, namely a graviton, a dilaton and a Kalb-Ramond field. In
the R-R sector, instead, we have the following massless states in the picture
$(-\frac{1}{2}, -\frac{1}{2})$:
\begin{equation}
u_A (k) {\tilde{u}}_{B} (k) |A  \rangle_{-\frac{1}{2}} |{\widetilde {B }}
\rangle_{-\frac{1}{2}} .
\label{rrsta7}
\end{equation}
Since the terms containing $G$ and ${\tilde{G}}$ in the second
equation in (\ref{gso}) act as the identity, the projector
$P_{\rm R-R}$ imposes the existence of two kinds of R-R
$(p+1)$-potentials for any value of $p$ ($C_{p+1}$ and
$\bar C_{p+1}$) characterized
respectively by:
\begin{equation}
u_A \left(\frac{1+ \Gamma_{11}}{2} \right)^{A}_{\,\,B} =0
\hspace{.5cm}; \hspace{.5cm} {\tilde{u}}_A \left(\frac{1 \pm
\Gamma_{11}}{2} \right)^{A}_{\,\,B} =0
\label{pro74}
\end{equation}
and by
\begin{equation}
u_A \left(\frac{1- \Gamma_{11}}{2} \right)^{A}_{\,\,B} =0
\hspace{.5cm}; \hspace{.5cm} {\tilde{u}}_A \left(\frac{1 \mp
\Gamma_{11}}{2} \right)^{A}_{\,\,B} =0
\label{pro75}
\end{equation}
where the upper [lower] sign corresponds to 0B [0A]. The doubling
of the R-R potentials implies the existence of two kinds of  branes
that are  charged with respect to
both the potentials. We follow the convention of denoting by $p$ and $p'$
respectively branes having equal or opposite charges with
respect to the two $(p+1)$ R-R potentials.
The $p$ and $p'$-branes
are called respectively
electric and magnetic branes \cite{KT9811}. In the case of a D3-brane
the two potentials $C_4$ and ${\tilde{C}}_4$ have field strenghts
$F_5 $ and ${\tilde{F}}_5$ that are respectively self-dual, $F_5 = {}^{*} F_5$,
and antiself-dual, ${\tilde{F}}_5 = - {}^*
{\tilde{F}}_5$,  as follows from Eqs. (\ref{pro74}) and (\ref{pro75}).
This means that, if we
take the linear combinations:
\begin{equation}
\label{elmag}
(C_4)^\pm=\frac{1}{{\sqrt 2}}(C_4\pm\bar C_4)
\end{equation}
one can see that the Hodge duality transforms each field strength into the other according to the relation $\,^*F_{5}^{\pm}=F_{5}^{\mp}$.
Therefore, while the D3-brane of Type
IIB is naturally dyonic, in Type 0B the dyonic D3-brane is
constructed as a superposition of an equal number of electric
and magnetic D3-branes.

The existence of two different kinds of branes in Type 0 theories
(the $p$-brane and the $p'$-brane) implies the one of four distinct
kinds of open strings: those stretching between two $p$ or two
$p'$-branes (denoted by $pp$ and $p'p'$) and those of mixed type ($pp'$ and
$p'p$).  This means that the most general Chan-Paton factor
$\lambda$ in the expression of the open string states has the following form:
\begin{eqnarray}
\lambda\equiv\left( \begin{array}{ll}
                     pp & pp'\\
                     p'p &p'p'
                     \end{array}
                     \right) .
\label{cpf}
\end{eqnarray}
Open/closed string duality makes the following spin structure
correspondence to hold \cite{KT9811}:
\begin{eqnarray}
\begin{array}{lll}
       {\rm Interactions} &
{\rm Closed\,\,states}  & {\rm Open\,\,states}\\
       pp\,\,\,\,p'p'& {\rm NS-NS}\,\,\,\,\,\,
&{\rm NS}\\
pp\,\,\,\,p'p'&{\rm R}\,-\,{\rm R} &{\rm NS}(-1)^F\\
pp'\,\,\,\,p'p &{\rm NS-NS}(-1)^F\,\,\,\,\,\,&{\rm R}\\
pp'\,\,\,\,p'p &{\rm R-R}(-1)^F\,\,\,\,\,\,&{\rm R}(-1)^F
\end{array}
\label{schema}
\end{eqnarray}
From this scheme, it is  easy to see that the spectrum of $pp$ and
$p'p'$ strings contains only the NS and NS$(-1)^F$ sectors
\cite{KT9811} \cite{BFL9904} whose massless excitations are the bosons of
the gauge theory. In the case of $D3$-branes
one has:
\begin{eqnarray}
&& A^{\alpha} \equiv \left(\begin{array}{ll}
                     pp & 0\\
                      0&p'p'
                     \end{array}
                     \right)\,\otimes \psi^{\alpha}_{-1/2}|0\rangle_{-1} \qquad
                     \alpha=0, \dots, 3 \label{boson} \\
&& \phi^{i} \equiv \left(\begin{array}{ll}
                     pp & 0\\
                      0&p'p'
                     \end{array}
                     \right)\,\otimes \psi^{i }_{-1/2}|0\rangle_{-1} \qquad
                     i=4, \dots, 9
\label{boson1}
\end{eqnarray}
$A^{\alpha}$ corresponds to the gauge vector of the gauge theory, while
the $\phi^{i}$'s represent six adjoint scalars. On the other hand $p
p'$ strings have only the R spectrum \cite{KT9811} \cite{BFL9904} which
provides fermions to the gauge theory supported by the
branes. The lowest excitation of these strings are:
\begin{eqnarray}
\psi^{A} \equiv \left(\begin{array}{ll}
                     0 & pp'\\
                      p'p&0
                     \end{array}
                     \right)\, \otimes |A\rangle_{-\frac{1}{2}} .
\label{ferm}
\end{eqnarray}
being $|A\rangle_{-\frac{1}{2}}$ a Majorana-Weyl spinor of the
ten-dimensional Lorentz group. The interaction between two branes
of the same kind is \cite{KT9811}:
\begin{eqnarray}
Z^{o}_{pp}&=&2\int \frac{d \tau}{2 \tau} \mbox{Tr}_{\rm NS}
\left[e^{-2\pi \tau L_0}
 (-1)^{G_{bc}}\left(
\frac{(-1)^{G} + (-1)^F }{2} \right)
\right]\nonumber \\
&=& V_{p+1}(8\pi^2\alpha')^{-\frac{p+1}{2}}\int
\frac{d\tau}{\tau^{\frac{p+3}{2} }}
e^{-\frac{y^2\tau}{2\pi\alpha'}} \frac{1}{2}\left[\left(
\frac{f_3(k)}{f_1(k)}\right)^8-\left(
\frac{f_4(k)}{f_1(k)}\right)^8\right]
 \label{free0pp}
\end{eqnarray}
where $o$ stands for open, $y$ is the distance between the branes
and $G_{bc}$ is the ghost number operator. The interaction between
a $p$ and  a $p'$-brane is obtained by computing the trace in the
R-sector \cite{KT9811}:
\begin{eqnarray}
Z^{o}_{pp'} &=&2\int \frac{d \tau}{2 \tau} \mbox{Tr}_{\rm R}
\left[e^{-2\pi \tau L_0}
 (-1)^{G_{bc}}\left(
\frac{(-1)^{G} + (-1)^F }{2} \right)
\right] \nonumber\\
&=&- V_{p+1}(8\pi^2\alpha')^{-\frac{p+1}{2}}\int \frac{d\tau}{\tau^{\frac{p+3}{2}}} e^{-\frac{y^2\tau}{2\pi\alpha'}}
\frac{1}{2}\left( \frac{f_2(k)}{f_1(k)}\right)^8
\label{free0pp'}
\end{eqnarray}
The explicit expressions for the functions $f_{i} (k)$ can be
found in App. A of Ref.~\cite{LMP}. Eqs. (\ref{free0pp}) and
(\ref{free0pp'}) can be written in a more compact form by introducing,
in the trace of the free energy, the following projector:
\begin{eqnarray}
P_{(-1)^{F_s}} =\frac{1+(-1)^{F_s}}{2}
\label{pfs}
\end{eqnarray}
that, as we have mentioned above, eliminates all fermionic states from
the spectrum since $F_s$ is the space-time fermion number operator. The
introduction of this operator is quite natural if one regards
Type 0 string theory as the orbifold Type IIB$/(-1)^{F_s}$ \cite{NS9902}.
From this point of view, the spectrum of
the  physical closed string states, written in Eqs.
(\ref{70as}) and (\ref{70bs}), is made of the untwisted and
twisted sectors of this orbifold.
Of course the untwisted spectrum coincides with the bosonic states
of Type II theories. The twisted sector can be more easily determined using
the Green-Schwarz formalism, rather than the NS-R one, due to the
simple action of $(-1)^{F_{s}}$ on the
space-time fermion
coordinates $S^{Aa}$. Here $A=1,2$ and $a$ labels
the two spinor representations of the light-cone Lorentz group $SO(8)$, namely
it is either an
$\bf 8_{s}$ or $\bf 8_{c}$ index. In the twisted sector the boundary
conditions on these coordinates are antiperiodic rather than periodic.
Hence, the Fourier
expansion for them  contains half-integer fermion modes. The lowest
level corresponds to a tachyon while the first one, corresponding to
the massless states, is given by:
\begin{eqnarray}
&& S^{a}_{-\frac{1}{2}} \tilde{S}^{b}_{-\frac{1}{2}} |0 \rangle | \otimes
 |\tilde{0} \rangle
\qquad \mbox{in Type 0B}\\
&& S^{a}_{-\frac{1}{2}} \tilde{S}^{\dot{b}}_{-\frac{1}{2}} |0 \rangle \otimes
 | \tilde{0} \rangle
\qquad \mbox{in Type 0A} .
\end{eqnarray}
These states provide the doubling of the R-R forms previously discussed.

The spectrum of  open strings can
be easily obtained in the NS-R superstring formalism. Since a stack of
$N$ dyonic D-branes of Type 0 contains $N$ $p$-branes and $N$ $p'$-branes, 
the massless open string states living on their
world-volume are the subset of the massless open string states living on the world-volume of
$2N$ D$p$-branes of Type II
theories, that are invariant under the action
of the operator  $(-1)^{F_s}$. However, in order to recover the
results given in Eqs. (\ref{boson}), (\ref{boson1}) and (\ref{ferm}),
we have to define a non trivial action of the space-time fermion
number operator on the Chan-Paton factors,
i.e. \cite{NS9902}~\cite{BFL9906}:
\begin{eqnarray}
(-1)^{F_s} \lambda_{ij} \equiv \left( \gamma_{(-1)^{F_s}} \right)_{ih}
\lambda_{hk}
\left( \gamma_{(-1)^{F_s}}^{-1}\right)_{kj}
\label{pfscp}
\end{eqnarray}
where
\begin{eqnarray}
\gamma_{(-1)^{F_s}}=\left(
\begin{array}{cc}
\mathbb{ I }_{N{\times} N} & 0 \\
0 & - \mathbb{ I }_{M{\times} M}
\end{array} \right)
\label{gpfs}
\end{eqnarray}
and $N$, $M$  denote the number of $p$ and  $p'$-branes respectively. The
requirement of invariance for the physical states imposes the
following constraints on the Chan-Paton factors:
\begin{eqnarray}
\lambda^{\rm (NS)}= \gamma_{(-1)^{F_s}} \lambda^{\rm (NS) }
\gamma_{(-1)^{F_s}}^{-1} \qquad
\lambda^{\rm (R)}= -\gamma_{(-1)^{F_s}} \lambda^{\rm (R) }
\gamma_{(-1)^{F_s}}^{-1}
\label{ccp}
\end{eqnarray}
where the minus sign is due to the action of $F_s$ on the space-time
fermion $|A \rangle$. It is easy to see that the previous equations
are satisfied  by the matrices given in
Eqs. (\ref{boson}), (\ref{boson1}) and (\ref{ferm}).
The free-energy is now written as:
\begin{eqnarray}
&&Z^{o}=2\int \frac{d \tau}{2 \tau} \mbox{Tr}_{\rm NS-R}
\left[e^{-2\pi \tau L_0} P_{\rm GSO}\left(\frac{1+(-1)^{F_s}}{2}\right)\right]\nonumber \\
&&= \frac{1}{2} {\rm Tr}\left[\,\mathbb{ I }\,\right]^2 V_{p+1}(8\pi^2\alpha')^{-\frac{p+1}{2}}\int \frac{d\tau}{\tau^{\frac{p+3}{2}}} e^{-\frac{y^2\tau}{2\pi\alpha'}}
\frac{1}{2}\left[ \left(
\frac{f_3(k)}{f_1(k)}\right)^8-\left(
\frac{f_4(k)}{f_1(k)}\right)^8-\left( \frac{f_2(k)}{f_1(k)}\right)^8\right]\nonumber\\
&& + \frac{1}{2} {\rm Tr}\left[\gamma_{(-1)^{F_s} }\right]^2\!
V_{p+1}(8\pi^2\alpha')^{-\frac{p+1}{2}}\!\int \!
\frac{d\tau}{\tau^{\frac{p+3}{2}}} e^{-\frac{y^2\tau}{2\pi\alpha'}}
\frac{1}{2}\left[ \left(
\frac{f_3(k)}{f_1(k)}\right)^8\!\!-\left(
\frac{f_4(k)}{f_1(k)}\right)^8\!\!+
\left( \frac{f_2(k)}{f_1(k)}\right)^8\right]\nonumber\\
&&\equiv Z_{pp}^{o} + 2 Z_{pp'}^{o} + Z_{p'p'}^{o} \label{cfre}
\end{eqnarray}
Since the traces of the Chan-Paton factors are given by
\begin{eqnarray}
{\rm Tr}\left[\,\mathbb{ I }\,\right]^2=(N+M)^2 \qquad  {\rm Tr}\left[\gamma_{(-1)^{F_s} }\right]^2=(N-M)^2 ,
\end{eqnarray}
we can rewrite the previous equation as follows:
\[
Z^{o} =  V_{p+1}(8\pi^2\alpha')^{-\frac{p+1}{2}}\!\int \!
\frac{d\tau}{\tau^{\frac{p+3}{2}}}
e^{-\frac{y^2\tau}{2\pi\alpha'}} 
\]
\begin{equation}
{\times} \left\{  \frac{N^2 + M^2}{2} \left[   \left(
\frac{f_3(k)}{f_1(k)}\right)^8-\left(
\frac{f_4(k)}{f_1(k)}\right)^8 \right] -{MN} \left(
\frac{f_2(k)}{f_1(k)}\right)^8 \right\} .
\label{zop89}
\end{equation}
We see that Eqs. (\ref{free0pp}) and (\ref{free0pp'}) are obtained by putting
respectively $M=0$, $N=1$ and $N=M=1$ in Eq. (\ref{zop89}),
while by taking $N=M$ we get the interaction among composite objects
made of an equal number
of D$p$ and D$p'$-branes.

To conclude, let us observe that the world-volume of a dyonic D3-brane 
configuration (corresponding to the case $N=M$) supports a
$U(N){\times} U(N)$ gauge theory with six adjoint scalars for each
gauge factor and four Weyl fermions in the bi-fundamental
representation of the gauge group $(N\,,\,\bar{N})$ and
$(\bar{N}\,,\,N)$. The number of bosonic degrees of freedom of the
open strings attached on the $N$ dyonic branes is $8N^2
{\times}2$, and coincides with the number of the fermionic ones.
Therefore, the gauge theory supported by these bound states, even
if non-supersymmetric, exhibits a Bose-Fermi degeneracy. From
this point of view, the interaction between two dyonic branes
vanishes because of a perfect cancellation between the
contribution of bosonic and fermionic degrees of freedom, ensuring
the stability of the configuration. The Bose-Fermi degeneracy
turns out to be an essential ingredient for making the 
gauge/gravity correspondence to hold. We will come back on this point
later.

\section{Orientifold $\Omega' I_6 (-1)^{F_L}$ of Type $0B$}
\label{orienti}
In this section we study
Type 0 string theory on the orientifold $\Omega' I_6 (-1)^{F_L}$,
where $\Omega'$ is the world-sheet parity operator, $I_6$ is the
inversion on the six space-time coordinates labelled by $i=4, \dots,
9$, i.e:
\begin{eqnarray}
I_6:\,\, (x^4, \dots, x^9) \rightarrow   (-x^4,\dots, -x^9)
\end{eqnarray}
and ${F_L}$ is the space-time fermion number operator in the left sector.

The orientifold fixed plane is, by definition,  the set of points left
invariant by the combined  action of $\Omega'$ and $I_6$. In our case
such a plane is located at $x^4=\dots=x^9=0$.

The action of the world-sheet parity operator ${\Omega'}$ in the open
string sector is:
\begin{eqnarray}
{\Omega'} \alpha_{m} {\Omega'}^{-1} = \pm e^{i \pi m} \alpha_m
\qquad {\Omega'}   \psi_r {\Omega'}^{-1} = \pm e^{i \pi r} \psi_r
\label{omegaos}
\end{eqnarray}
for integer and half-integer $r$ and the signs $\pm$ refer
respectively to NN and DD boundary conditions.
The $\Omega'$ action on the ghost and superghost oscillators is instead
given by:
\begin{eqnarray}
{\Omega'} b_{n} {\Omega'}^{-1} =  e^{i \pi n} b_n
\qquad {\Omega'}   \beta_r {\Omega'}^{-1} =  e^{i \pi r} \beta_r
\label{omegagh}
\end{eqnarray}
and by analogous expressions for the $c$-ghost and the $\gamma$-superghost.
Finally, its action on
the NS and R vacua is
\begin{eqnarray}
{\Omega'} |0 \rangle_{-1} = -i |0 \rangle_{-1} \label{omegavac}
\end{eqnarray}
\begin{eqnarray}
{\Omega'} |A \rangle_{-\frac{1}{2}} =
- |A \rangle_{-\frac{1}{2}} \qquad {\Omega'} |A \rangle_{-\frac{1}{2}} = -
\Gamma^{p+1} \dots \Gamma^{9} |A \rangle_{-\frac{1}{2}}
\label{ramo23}
\end{eqnarray}
where the first equation in (\ref{ramo23}) must be used in the case of
NN boundary conditions, while the second is valid for DD boundary
conditions along the directions  $\{p+1, \dots, 9 \}$.

The action of $\Omega'$ in the closed string sector is\footnote{ In
  Refs. \cite{FGLS}, \cite{tesia} and \cite{LR9905} a
different definition of the world-sheet parity operator
is used where the action of $\Omega'$ on the R-R vacuum is not as in
  Eq. (\ref{omegavuo}).} :
 \begin{eqnarray}
&& \hspace{-1cm} \Omega' \alpha^{\mu}_n \Omega^{'-1}  =  \tilde{\alpha}^{\mu}_n \qquad
\Omega' \tilde{\alpha}^{\mu}_n \Omega^{'-1}  =  \alpha^{\mu}_n \qquad
\Omega' \psi^{\mu}_r \Omega^{'-1} =  \tilde{\psi}^{\mu}_r \qquad
\Omega' \tilde{\psi}^{\mu}_r \Omega^{'-1}  =   \psi^{\mu}_r \nonumber \\
&& \hspace{-1cm} \Omega' b_n \Omega^{'-1}  =  \tilde{b}_n \qquad
\Omega' c_n \Omega^{'-1}  =  \tilde{c}_n \qquad
\Omega' \beta_r \Omega^{'-1} =  \tilde{\beta}_r \qquad
\Omega' \gamma_r \Omega^{'-1}  =   \tilde{\gamma}_r
\label{ome65}
\end{eqnarray}
and
\begin{eqnarray}
\Omega' \left(|0 \rangle_{-1} \otimes |\tilde{0} \rangle_{-1} \right)& = &
- |0 \rangle_{-1} \otimes |\tilde{0} \rangle_{-1} \label{ome66} \\
\Omega' \left(|A \rangle_{-\frac{1}{2}} \otimes | \tilde{B}
\rangle_{\frac{1}{2}}\right) & = & \left(
\Gamma^{11} \right)^{B}_{\,\,\,D} | D \rangle_{-\frac{1}{2}} \otimes | \tilde{A}
\rangle_{-\frac{1}{2}} .
\label{omegavuo}
\end{eqnarray}
Finally, let us notice that the reason why the orientifold
projector contains the term $(-1)^{F_L}$ with the space-time
fermion number operator
of the left sector is that, in general, the operator $\Omega'
I_{2n}$ squares to unity only for $n$ even. In fact $I^2_{2n}$
represents a $2\pi$ rotation in $n$ planes and, for n odd, is
equal to $(-1)^{F_s}=(-1)^{F_L+F_R}$. Therefore:
\begin{eqnarray}
\left[\Omega'I_{2n}
(-1)^{F_L}\right]^2= I_{2n}^2(-1)^{F_L+F_R}=\mathbb{I}\qquad n=1,3
\end{eqnarray}
where we have used the fact that $\Omega' (-1)^{F_L} \Omega'^{-1}
= (-1)^{F_R}$  and $\Omega'^{2}=1$. The introduction of
$(-1)^{F_{L}}$ is also consistent with the general property that
making the  orientifold projection $\Omega' I_n$ is equivalent to
performing  $n$ T dualities on the unoriented theory.  T-duality
transforms $\Omega'$ into $\Omega' I_n (-1)^{F_L}$ where the last
operator is again present only if one rotates fermions in an odd
number of planes (i.e for $I_2$ and $I_6$).

\subsection{Open and closed string spectrum}
\label{opeclo}

Let us determine the spectrum of the massless open string states
attached to $N$ D3-branes at the orientifold plane. The
generic massless open string state in Type 0 is given by:
\begin{eqnarray}
A^{\alpha}\equiv\lambda_A \psi^{\alpha}_{-1/2} |0, k\rangle~~,~~
\phi^{i}\equiv\lambda_\phi \psi^{i}_{-1/2} |0, k\rangle~~,~~ \psi
\equiv \lambda_\psi |s_0\dots s_4\rangle
\end{eqnarray}
being $\lambda$ a $2N{\times} 2N$ matrix, $\alpha=0,\dots, 3 $ and $i=4,
\dots, 9$. By
imposing the invariance under the space-time fermion number operator
given in Eqs. (\ref{pfs}) and (\ref{gpfs}), we obtain for the
bosonic Chan-Paton factors the diagonal structure given in Eqs.
(\ref{boson}) and (\ref{boson1}) and for the fermionic ones the
off-diagonal structure in Eq. (\ref{ferm}). On these states we have
then to impose the orientifold projection and select only those states
that are invariant under the action of $\Omega'I_6$. In the
NS-sector we have:
\begin{eqnarray}
\Omega'I_6 \,\,\,  \psi^{\alpha}_{-1/2} |0, k\rangle\rightarrow
- \psi^{\alpha}_{-1/2} |0, k\rangle~~,~~
 \Omega'I_6 \,\,\,\psi^{i}_{-1/2} |0, k\rangle\rightarrow
-\psi^{i}_{-1/2} |0, k\rangle
\end{eqnarray}
and therefore the invariant states satisfy the constraint:
\begin{eqnarray}
\gamma_{\Omega'I_6}\lambda^T_{A, \phi}\gamma_{\Omega'I_6}^{-1} =
-\lambda_{A, \phi} .
\label{bos12}
\end{eqnarray}
In the R-sector, instead, we have to determine how the reflection
operator acts on the spinor state. By observing that a reflection in a
plane corresponds to
a rotation of an angle $\pi$ in the plane, we can write:
\begin{eqnarray}
I_6\equiv {\rm e}^{\pm i\pi(S^{45}+S^{67}+S^{89})},
\end{eqnarray}
being $S^{ij}$ the zero modes of the Lorentz group generators, i.e:
\begin{eqnarray}
S^{ij}=-\frac{i}{2} [\psi^i_0,\,\psi^j_0]~~~,~~~ \sqrt{2}\psi_{0}^{i}
\equiv \Gamma^i .
\end{eqnarray}
By introducing the operators
\begin{eqnarray}
N_0=-\frac{\Gamma^0 \Gamma^1}{2}
~~~,~~~N_{i}=-i\frac{\Gamma^{2i}\Gamma^{2i+1}}{2}
\qquad\mbox{ with}\,\,\,i=1,\dots,4
\label{ni56}
\end{eqnarray}
it is straightforward to verify that $N_2\equiv S^{45}$,
$N_3\equiv S^{67}$ and $N_4\equiv S^{89}$ .
In conclusion we get:
\begin{eqnarray}
I_6|s_0\dots s_4\rangle=e^{\pm i\pi( s_2 +s_3+s_4)}|s_0 \dots s_4\rangle=
 \prod_{i=1}^{3} ( \pm 2i N_i)|s_0 \dots s_4\rangle =
\pm \Gamma^4\dots \Gamma^9|s_0\dots s_4\rangle
\label{act12}
\end{eqnarray}
where we have taken into account that the state $|s_0 \dots s_4\rangle
$  is an eigenstate of the operator $N_i$ with eigenvalue
$s_i =\pm 1/2$. By choosing in Eq. (\ref{act12}) the minus sign we
obtain  the result
given in Ref. \cite{D9804}, while by
choosing the plus sign we find  agreement with Ref. \cite{GP9601}. We
follow the latter convention and therefore in the R-sector we
write:
\begin{eqnarray}
\Omega'I_6|s_0\dots s_4\rangle= |s_0\dots s_4\rangle \Longrightarrow
\gamma_{\Omega'I_6}
\lambda^T_{\psi}\gamma_{\Omega'I_6}^{-1} =\lambda_{\psi}
\label{ferm12}
\end{eqnarray}
In the previous equation we should also take into account the action
of the operator $(-1)^{F_L}$. This is irrelevant or gives an
extra minus sign in the R-sector  depending whether the open string is
considered respectively the right or left sector of the closed
string. However, it is simple to check that this sign ambiguity is
completely irrelevant in determining the spectrum of the massless
states.

In the last part of this section we determine the orientifold
action on the Chan-Paton factors. First we observe that the
Chan-Paton factors have to be $2N{\times} 2N$ matrices in order to
take into account their images under $\Omega'I_6$. Furthermore,
following Ref. \cite{GP9601}, they have to satisfy the constraint
$\gamma_{\Omega'I_6}=\pm \gamma_{\Omega'I_6}^T$ that implies
\begin{eqnarray}
 \gamma_{\Omega'I_6}=\left( \begin{array}{cc}
                            0            &\mathbb{I}_{N{\times} N}\\
                           \pm\mathbb{I}_{N{\times} N} &  0
                           \end{array}
                      \right) .
\label{g0i6}
\end{eqnarray}
Substituting Eq. (\ref{g0i6}) in  Eqs. (\ref{bos12}) and
(\ref{ferm12}), one gets for the bosonic and fermionic Chan-Paton
factors:
\begin{equation}
\lambda_{A, \phi} = \left( \begin{array}{cc}
                      A    & 0 \\
                      0 &  - A^T   \end{array}   \right)
\,\,\,\,\,\,\,\,\,\,\,\,\,\,\,
\lambda_{\psi} = \left( \begin{array}{cc}
                      0    & B \\
                      \pm B^{*} &  0   \end{array}   \right)
\label{equa46}
\end{equation}
where in the last expression we have implemented the hermiticity
of the Chan-Paton factor and the matrix B can be chosen to be
either symmetric or antisymmetric depending on how the sign in
Eq. (\ref{g0i6}) is chosen. The number of bosonic degrees of
freedom is $8N^2$ which corresponds to one gauge boson and six real
scalars transforming according to the adjoint representation of
$SU(N)$. In the fermionic sector one has $8N^2\pm 8N$
corresponding to four Dirac fermions in the two-index symmetric
($+$) or antisymmetric ($-$) representation. Notice that the
spectrum does not satisfy the Bose-Fermi degeneracy condition that
holds in Type $0$ theory. In this case such a degeneracy is
present only in the large $N$ limit.

Moreover the spectrum of this theory has the same bosonic content
as ${\cal N}=4$ SYM. This is an example of planar equivalence
\cite{ASV} between a supersymmetric model, the ${\cal N}=4$ SYM,
which plays the role of the {\it parent} theory and a 
non-supersymmetric one, that is the orientifold $\Omega' I_6
(-1)^{F_L}$ of Type $0B$, which is the {\it daughter} theory, the
two being equivalent in the large $N$ limit. In section
\ref{1lof}, using string techniques, we will explicitly see that
in this limit the two theories have the same $\beta$-function.

Let us consider the closed string spectrum. Since $\Omega'$ leaves
invariant the metric and the dilaton, while changes sign to the
Kalb-Ramond field it is easy to see that in the  NS-NS sector the
orientifold projection selects the following states
\begin{equation}
\phi,\,\,\,g_{\alpha\beta},\,\,\,g_{ij},\,\,\,B_{i\alpha}
\,\,\,\,\,\,\,\,\,{\rm with}\,\,\,\,\,\,\alpha, \beta =0,\dots,3
\,\,\,\,\,\,i,j=4,\dots,9
\label{spnsns}
\end{equation}
where $\phi$, $g$ and $B$ are respectively the dilaton, graviton and
Kalb-Ramond fields. In the R-R sector the states which are even under
the orientifold projection are
\begin{equation}
{\rm (R+,R+)} \,\,\,\,\,\,\rightarrow\,\,\,\,\,\,
C_0,\,\,\,\,
C_{\alpha i},\,\,\,\,
C_{0123},\,\,\,\,
C_{\alpha \beta ij },\,\,\,\,
C_{ijhk},\,\,\,\,
\label{sprr++}
\end{equation}
\begin{equation}
{\rm (R-,R-)} \,\,\,\,\,\,\rightarrow\,\,\,\,\,\,
\bar C_{\alpha \beta},\,\,\,\,
\bar C_{ij},\,\,\,\,
\bar C_{\alpha \beta \gamma  i},\,\,\,\,
\bar C_{\alpha ijk },\,\,\,\,
\label{sprr--}
\end{equation}
The previous results follow from the fact that, because of
 Eq. (\ref{omegavuo}),  in the sector ${\rm
  (R+,R+)}$ where $\Gamma_{11}=1$, $\Omega '$ leaves $C_2$ invariant
and changes the sign of $C_0$ and $C_4$. In the sector  ${\rm
(R-,R-)}$ one has instead $\Gamma_{11} =-1$ and therefore
$\Omega'$  leaves $\bar C_{0}$ and $\bar C_{4}$ invariant and
changes the sign of $\bar C_{2}$. Notice that the R-R 5-form field
strength surviving the orientifold projection is the self-dual one
($dC_4=\,^*dC_4$), while the anti-self dual one ($d\bar
C_4=-\,^*d\bar C_4$) is projected out. The twisted sector is
simply made of open strings with NN boundary conditions in the
directions $\{0,\dots,3\}$ and DD boundary conditions in the
directions $\{4,\dots,9\}$.

\subsection{One-loop vacuum amplitude}

In this section we compute the interaction between two stacks of
$N$ D3-branes in the orientifold $\Omega' I_6$,
 the action
of the operator $(-1)^{F_L}$ being irrelevant in the open string
calculation, as previously discussed.

The one-loop open string gets two contributions, the annulus
$Z_{e}$ and the Moebius strip $Z_{\Omega'I_6}$:
\begin{eqnarray}
Z^{o}&\equiv &  Z_{e}^o + Z_{\Omega'I_6}^o\nonumber\\
&=&\int_{0}^{\infty} \frac{d \tau}{\tau} Tr_{{\rm
    NS-R}} \left[ \frac{e+\Omega'I_6}{2}\,
\frac{1+  (-1)^{F_s}}{2}\,(-1)^{G_{bc}}\,
\frac{(-1)^{G} + (-1)^F }{2}  { e}^{- 2 \pi \tau
L_0} \right] \label{Z1}
\end{eqnarray}
The annulus contribution is equal to the one in Eq. (\ref{zop89})
with $M=N$ and with an extra factor $1/2$ due to the orientifold
projection. For $p=3$ we get:
\begin{equation}
Z^{o}_e = N^2 \frac{V_{4}}{(8\pi^2\alpha')^{2}}\!\int \!
\frac{d\tau}{2\tau^{3}}
\left[   \left( \frac{f_3(k)}{f_1(k)}\right)^8\!\!-\left(
\frac{f_4(k)}{f_1(k)}\right)^8\!\!  - \left(
\frac{f_2(k)}{f_1(k)}\right)^8 \right] . \label{182} 
\end{equation}
In particular it vanishes because of the abstruse identity. The
contribution of the Moebius strip, corresponding to the insertion of
$\Omega' I_6$ in the trace, is instead non-trivial. Let us
first compute, for such a term, the trace over the Chan-Paton
factors. By fixing the normalization as $\langle hk|nm\rangle
=\delta_{kn}\delta_{hm}$, one finds:
\begin{eqnarray}
&&{\rm Tr}^{\rm C.P.} \left[ \langle hk| \Omega'I_6|ij\rangle\right]=
{\rm Tr}\left[\gamma_{\Omega'I_6}^{-1}
\gamma_{\Omega'I_6}^{T}\right]\nonumber\\
&&{\rm Tr}^{\rm C.P.} \left[ \langle hk| \Omega'I_6 (-1)^{F_s}|ij
\rangle\right]=
{\rm Tr}\left[\gamma_{\Omega'I_6}^{-1} \gamma^{-1}_{(-1)^{F_s}}
\gamma_{\Omega'I_6}^{T}
\gamma^{T}_{(-1)^{F_s}}\right] .
\label{cp03}
\end{eqnarray}
Furthermore, from the explicit form of the matrices introduced in the
last expression and given in Eqs. (\ref{gpfs}) and (\ref{g0i6}), it is
straightforward to check that
\begin{eqnarray}
{\rm Tr}\left[\gamma_{\Omega'I_6}^{-1} \gamma^{-1}_{(-1)^{F_s}}
\gamma_{\Omega'I_6}^{T}
\gamma^{T}_{(-1)^{F_s}}\right]= - {\rm
Tr}\left[\gamma_{\Omega'I_6}^{-1}
\gamma_{\Omega'I_6}^{T}\right] .
\label{cp103}
\end{eqnarray}
This identity implies that the NS contribution to the free energy vanishes.

A non-vanishing contribution comes from the R sector, where the
trace over the non-zero modes ($n.z.m.$) gives
\begin{eqnarray}
&&{\rm Tr}_R^{n.z.m.}\left[e^{-2\pi\tau
N_\psi}\Omega'I_6(-1)^G\right] = \frac{(ik)^{-2/3}}{2^4}
f_2^8(ik)\label{tro03}\\
&&{\rm Tr}_R^{n.z.m.}\left[e^{-2\pi\tau
N_\psi}\Omega'I_6(-1)^F\right] = (ik)^{-2/3}
f_1^8(ik) ,
\label{trof03}
\end{eqnarray}
while the trace over the zero modes ($z.m.$) is given by:
\begin{eqnarray}
&&{\rm Tr}_R^{z.m.}\left[\Omega'I_6(-1)^{G_0}\right] =
{\rm Tr}_R^{z.m.}\left[(-1)^{G_0}\right]
= 2^4\label{trozm03}\\
&&{\rm Tr}_R^{z.m.}\left[\Omega'I_6(-1)^{F_0}\right] =
{\rm Tr}_R^{z.m.}\left[(-1)^{F_0}\right]=0 .
\label{trofzm03}
\end{eqnarray}
By inserting Eqs. (\ref{cp03}), (\ref{cp103})  and
(\ref{tro03} $\div$ \ref{trofzm03}) in the term with  $\Omega ' I_6$ in
Eq. (\ref{Z1}), we get:
\begin{eqnarray}
Z_{\Omega'I_6}^{o}=-\frac{V_{4}}{4 (8\pi^2\alpha')^{2}}\, {\rm
Tr} \left[\gamma_{\Omega'I_6}^T \gamma_{\Omega'I_6}^{-1}\right]
\int_{0}^{\infty} \frac{d\tau}{\tau^3}
\left(\frac{
f_2(ik)}{f_1(ik)}\right)^8 \label{freeom03}
\end{eqnarray}
where we should use that ${\rm Tr} \left[\gamma_{\Omega'I_6}^T
\gamma_{\Omega'I_6}^{-1}\right]=\pm 2N$. Notice that, because of
the Moebius strip contribution, the interaction between two D$p$
branes in this orientifold, is non vanishing.

The previous equation, together with the third term of Eq.
(\ref{182}), gives the total fermionic contribution to the
free-energy which at the massless level reduces to:
\begin{eqnarray}
Z^{o}({\rm femionic\,massless})=-(8N^2\pm 8N) \frac{V_{4}}{
(8\pi^2\alpha')^{2}}\,  \int_{0}^{\infty} \frac{d\tau}{\tau^3}.
\label{dof4}
\end{eqnarray}
 As usual, the factor
$(8N^2\pm 8N)$ in front of the previous expression counts the
number of the fermionic degrees of freedom of the world-volume
gauge theory, which indeed agrees with the counting of the
previous subsection. As already noticed, we do not have the same
number of bosonic and fermionic degrees of freedom propagating in
the loop and, in particular, the additional $\pm 8N$ fermionic
term  comes from the Moebius strip, which therefore is responsable
of spoiling the Bose-Fermi degeneracy of the theory \cite{BFL9906}.
This contribution is subleading in the large $N$ limit.

Notice that Eq. (\ref{freeom03}), apart from the Chan-Paton
factors and the substitution $k\rightarrow ik$, is 1/2 of
the free energy describing, in the R-sector, the interaction
between two D3-branes in Type IIB string theory. It is, by the
way, also equal, with the previous substitutions, to the
correspondent expression in Type 0 theory, given in Eq.
(\ref{free0pp'}).

\subsection{One-loop vacuum amplitude with
an external field}
\label{1lof}

Let us consider, in the open channel, the interaction between a
D3-brane dressed with a constant
$SU(N)$ gauge field and a stack of $N$ undressed D3-branes.
The gauge field is chosen to have only the entries $\hat{F}_{01}=2\pi\alpha'\hat{f}$ and  $\hat{F}_{23}=2\pi\alpha'\hat{g}$
different from zero.
The presence of the external field modifies Eq. (\ref{freeom03}) as
follows:
\begin{eqnarray}
Z^{o}(F)_{\Omega'I_6}&=&
\frac{2\,{\mbox Tr}\left[\gamma_{\Omega'I_6}^T
\gamma_{\Omega'I_6}^{-1}\right]}{(8\pi^2\alpha')^2 } \int d^4x
\sqrt{-{\rm det} (\eta+\hat{F}) } \int_{0}^{\infty}
\frac{d\tau}{\tau}
{\rm e}^{-\frac{y^2\tau}{2\pi\alpha'}}
\sin\pi\nu_f \sin\pi\nu_g\nonumber\\
&&\frac{ f_2^4(ik)
\Theta_2\left(i\nu_f\tau|i\tau +1/2 \right) \Theta_2
\left(i\nu_g\tau|i\tau+1/2 \right) }{f_1^4(ik)
\Theta_1\left(i\nu_f\tau|i\tau+1/2 \right) \Theta_1
\left(i\nu_g\tau|i\tau+1/2 \right)}
\label{ZF03}
\end{eqnarray}
where \cite{LMP}:
\begin{eqnarray}
\nu_f=\frac{1}{2\pi i}\log\frac{1+2\pi\alpha'
\hat{f}}{1-2\pi\alpha' \hat{f}}, \qquad \nu_g=
\frac{1}{2\pi i}\log\frac{1-2\pi i\alpha'\hat{g}}{1+2\pi i
\alpha'\hat{g}}.
\label{nufg}
\end{eqnarray}
Notice that in the previous expression we should have put $y=0$ because all
branes are located at the orientifold point. However we keep $y \ne 0$
because it provides a natural infrared cutoff.
Eq. (\ref{ZF03}) describes the Moebius strip with the
boundary on the dressed brane. The trace over the
Chan-Paton factors gives $\pm 2$ counting the
dressed brane and its image under $\Omega'I_6$.
The overall factor 2, instead, is a consequence of the fact that
in the trace we
have to sum over two different but equivalent open string
configurations: the first one in which only the end-point of the string
parametrized by the world-sheet coordinate $\sigma=0$ is charged
under the gauge group, and the other one in which the gauge charge
is turned instead on the other end-point at $\sigma=\pi$.

We are now going to  compute the field theory limit of Eq.
(\ref{ZF03}) which gives the corresponding Euler-Heisenberg
effective action and the threshold corrections to the running
coupling constant.

The field theory limit can be performed by taking $\alpha'
\rightarrow 0$ and $\tau \rightarrow \infty$ in such a way that
the Schwinger parameter of the field theory
$\sigma=2\pi\alpha'\tau$ is kept fixed. Using in this limit the
equations (\ref{the12})$\div$(\ref{equa23}) given in Appendix, from
Eq. (\ref{ZF03}) we get:
\begin{equation}
Z^F\simeq\mp \frac{16}{(4\,\pi)^2 } \int d^4 x  \int_{0}^{\infty}
\frac{d\sigma}{\sigma}
{\rm e}^{-\frac{y^2}{(2 \pi \alpha')^2}\,\sigma}\,\,
\hat{f}\,\hat{g}\,\,\frac{ \cos(\sigma \hat{f}) \cosh(\sigma
\hat{g})}{ \sin(\sigma \hat{f}) \sinh(\sigma \hat{g})} .
\label{263}
\end{equation}
Finally, expanding Eq. (\ref{263}) up to the quadratic order in
the gauge fields yields:
\begin{equation}
Z^F(F^2)\simeq\mp \frac{16}{(4\,\pi)^2 } \int d^4 x
\int_{1/\Lambda^2}^{\infty} \frac{d\sigma}{\sigma^3}
{\rm e}^{-\mu^2\,\sigma}\,
\left[ 1- \frac{1}{3}
\sigma^2 ( \hat{f}^2 \,  -\,    \hat{g}^2 )\right] ,
\end{equation}
where $\Lambda$ is an UV cut-off  and $\mu = \frac{y}{2 \pi \alpha'}$
is an IR one. By using
\begin{equation}
{\hat{f}}^2 - {\hat{g}}^2 =
- \frac{1}{4} F^{a}_{\mu \nu} F^{a \mu \nu},
\label{f298}
\end{equation}
one gets the following expression for the running coupling constant:
\begin{equation}
\frac{1}{g^2_{\rm YM} (\mu)}= \frac{1}{g^2_{\rm YM} (\Lambda )}
\mp \frac{1}{3\pi^2} \log
 \frac{\mu^2}{\Lambda^2} , 
\label{run78}
\end{equation}
where the tree diagram contribution has been introduced by hand.
Finally from Eq. (\ref{run78}) one reads the expected
$\beta$-function
\begin{eqnarray}
\beta(g_{YM})={\pm}\frac{g_{YM}^{3}}{(4\pi)^2} \frac{16}{3} .
\label{beta}
\end{eqnarray}
As already observed, in the planar limit $N\rightarrow\infty$ with
$\lambda=g_{\rm YM}^2 N$ fixed, the ratio $\beta(g_{\rm
YM})/g_{\rm YM}$ reduces to zero and coincides with the one of its
parent theory ${\cal N}=4$ SYM.

In the last part of this subsection, in order to explore the
validity of gauge/gravity correspondence in this non-supersymmetric model, 
we evaluate the threshold corrections to the
running coupling constant.

The starting point is again Eq. (\ref{ZF03}) that we now expand up
to the quadratic order in the gauge fields without performing any
field theory limit (more details on the calculation can be found
in the Appendix) obtaining $( k = {e}^{- \pi \tau} )$:
\begin{eqnarray}
\frac{1}{g^2_{\rm YM}} =\pm \frac{1}{(4\pi)^2} \int_{0}^{\infty}
\frac{d\tau}{\tau}
e^{-\frac{y^2\tau}{2\pi\alpha'} }
 \left(  \frac{f_2(ik)}{f_1(ik)}\right)^8 \left[\frac{1}{3\tau^2} +k
\frac{\partial}{\partial k} \log f_2^4(ik)\right] .
\label{gym03}
\end{eqnarray}
Moreover by performing  the field theory limit and keeping the
product $ \sigma = 2 \pi \alpha' \tau$ fixed, we get
the same expression of the running coupling constant obtained from the Euler-Heisenberg
action given in  Eq. (\ref{run78}).

It is also interesting to write the interaction given by
Eq. (\ref{ZF03}) in the closed string channel by performing the
modular transformation $\tau=1/4t$, as shown in Appendix:
\begin{eqnarray}
Z^{c}(F)_{\Omega'I_6}&=& \mp\frac{1}{(8\pi^2\alpha')^2 } \int d^4x
\sqrt{-{\rm det} (\eta+\hat{F}) } \int_{0}^{\infty} \frac{dt}{t^3}
{\rm e}^{-\frac{y^2}{2\pi\alpha't}}
\sin\pi\nu_f \sin\pi\nu_g\nonumber\\
&&\frac{ f_2^4(iq) \Theta_2\left(\frac{\nu_f}{2}|it +\frac{1}{2}
\right) \Theta_2 \left(\frac{\nu_g}{2}|it +\frac{1}{2} \right)
}{f_1^4(iq) \Theta_1\left(\frac{\nu_f}{2}|it +\frac{1}{2} \right)
\Theta_1 \left(\frac{\nu_g}{2}|it +\frac{1}{2} \right)} .
\label{ZF05}
\end{eqnarray}
Expanding the previous equation up to the second order in the
external field gives (with $q = {e}^{-\pi t}$):
\begin{eqnarray}
\frac{1}{g^2_{\rm YM}} =\pm \frac{1}{(4\pi)^2} \int_{0}^{\infty}
\frac{dt}{4t^3}
e^{-\frac{y^2}{2\pi\alpha't} }
\left(
\frac{f_2(iq)}{f_1(iq)}\right)^8 \left[\frac{4}{3} + \frac{1}{
\pi}\partial_t \log f_2^4(iq)\right] .
\label{gymc03}
\end{eqnarray}
Under the inverse modular transformation $t=1/4\tau$ this equation
perfectly reproduces the expression obtained in the open channel
(Eq. (\ref{gym03})), as one can easily check by using
Eqs. (\ref{f1f2}).

The field theory limit of the previous expression, realized as
$t\rightarrow \infty$ and $\alpha'\rightarrow 0$ with $s=2\pi
\alpha't$ fixed, gives a vanishing contribution to
Eq. (\ref{gymc03}). One could be
led to conclude that the gauge/gravity correspondence does not
work in this non-supersymmetric model. However, in the planar
limit ($N\rightarrow \infty$ and $g_{\rm YM}^2N$ fixed) the theory
has a vanishing $\beta$-function, as noticed after Eq.
(\ref{beta}). Therefore one can conclude that {\it the
gauge/gravity correspondence holds in the large $N$ limit}, where
the Moebius strip contribution is suppressed and  {\it the gauge
theory recovers the Bose-Fermi degeneracy in its spectrum}.

\section{Orbifolds of previous orientifolds}
\label{orbiorienti}

In this section we consider some orbifolds of the previous
orientifold theory and, within this framework, we analyze the
world-volume gauge theory living on $N$ fractional branes. Those
are branes fixed at the orbifold singularity and having a non
conformal gauge theory on their world-volume. The open strings
attached to them have Chan-Paton factors transforming according to
irreducible representations of the orbifold group and this
property makes them the most elementary solitonic objects in the
theory. We start by discussing the gauge theory living on $N$
fractional D3-branes in the orbifold $C^2/ \mathbb{Z}_2$. In the
planar limit this theory, which is not supersymmetric,  shows some
interesting common features with  ${\cal N}=2$ super Yang-Mills.

Then we will turn to the more interesting case of the
orbifold $C^3/(\mathbb{Z}_2\times \mathbb{Z}_2)$. Here the gauge
theory living on $N$ fractional branes  is the one recently
discussed by Armoni-Shifman-Veneziano, that for $N=3$ reduces to  QCD
with one flavour.

\subsection{Orbifold $C^2/\mathbb{Z}_2$ }

The $\mathbb{Z}_2$ group is characterized by two elements $(1,h)$,
being $1$ the identity element and $h$ such that $h^2=1$. We
choose $\mathbb{Z}_2$ to act on the coordinates $x^{m}$ with
$m=6,7,8,9$. Therefore, introducing the complex combinations
$\vec{z} = (z^1, z^2)$, where $z^1=x^6+ix^7$, $z^2=x^8+ix^9$, the
non trivial element of the orbifold group acts on them  as
follows:
\begin{eqnarray}
h:(z_1 \,,\,z_2)\rightarrow ( - z_1 \,,\, - z_2). \label{orba}
\end{eqnarray}
The gauge theory living on the fractional D3-branes is the
$\mathbb{Z}_2$ invariant subsector of the open string spectrum
introduced in Sect. \ref{opeclo}. In order to construct it explicitly one
also needs to know the action of the orbifold on the spinors
states:
\begin{eqnarray}
h:|s_0=1/2, s_1\, s_2\,s_3\, s_4\rangle \rightarrow e^{i \pi
(s_3+s_4)}|s_0=1/2, s_1\, s_2\,s_3\, s_4\rangle\,\,, \label{orbf}
\end{eqnarray}
where $s_i=\pm 1/2$ and $s_0$ has been fixed to $1/2$ by the
mass-shell condition. Using Eqs. (\ref{orba}) and (\ref{orbf}) it
is easy to see that the spectrum contains one $SU(N)$ gauge field,
two real scalars in the adjoint representation of the gauge group
and two Dirac fermions in the two-index symmetric (or
antisymmetric) representation. Notice that the spectrum has a {\it
common sector} \cite{ASV} with ${\cal N}=2$ SYM, namely the
bosonic one. However, because of the fermionic contributions
which  are different, the one-loop $\beta$-function of our
theory contains a subleading correction in $1/N$ with respect to
${\cal N}=2$ $\beta$-function:
\begin{eqnarray}
\beta(g_{YM})= \frac{g_{YM}^{3}}{(4 \pi)^2} \left[- \frac{11}{3} N
+ 2 \frac{N}{6} + 2 \frac{4}{3} \frac{N\pm2}{2}  \right] =
\frac{g^3_{YM}N}{(4\pi)^2} \left[ 2 \pm\frac{8}{3N}\right]
\label{betaz2}
\end{eqnarray}
In the  large $N$ limit the subleading term in $1/N$  is
suppressed and the two $\beta$-functions coincide.
This
circumstance signals the existence of a planar equivalence between
the two theories at one-loop  and suggests the possibility of an
extension of such equivalence at all perturbative orders.

The free energy, which describes the interaction between a stack
of $N$ undressed  D$3$ branes and a dressed one, is given by:
\[
Z =  \int_{0}^{\infty} \frac{d \tau}{\tau} Tr_{{\rm
    NS-R}} \left[ (\frac{1 +h}{2}) \left(\frac{e+\Omega'I_6}{2}
\right)\left( \frac{1+  (-1)^{F_s}}{2}\right) \right.
\]
\begin{equation}
{\times} \left.  (-1)^{G_{bc}}\left( \frac{(-1)^{G}
+ (-1)^F }{2} \right) { e}^{- 2 \pi \tau L_0} \right]
 \equiv  Z_{e}^o + Z_{\Omega'I_6}^o + Z_{he}^o + Z_{h\Omega'I_6}^o
\label{Z1orb}
\end{equation}
where the trace over the Chan-Paton factors has been understood
and  we have used the following notation:
\begin{eqnarray}
Z_{e}^o&\equiv&\left(Z_{e}^o+Z_{e(-1)^{F_s}}^o\right)/2\,\,\,\,\,\,\,\,\,\,\,\,\,\,\,\,\,\,\,
Z_{he}^o\equiv\left(Z_{he}^o+Z_{he(-1)^{F_s}}^o\right)/2 \nonumber\\
Z_{\Omega'I_6}^o&\equiv&\left(Z_{\Omega'I_6}^o+Z_{\Omega'I_6(-1)^{F_s}}^o\right)/2
\,\,\,\,\,\,\,\,\,\,\,\,\,\, Z_{h\Omega'I_6}^o\equiv
\left(Z_{h\Omega'I_6}^o+Z_{h\Omega'I_6(-1)^{F_s}}^o\right)/2\,.
\label{defs}
\end{eqnarray}
However notice that  the term $(-1)^{F_s}$ gives a non vanishing
contribution to the trace on the Chan-Paton factors, only if it
appears together with the projector $\Omega'I_6$, namely in the
terms appearing in the second line  of Eq. (\ref{defs}).

The first two terms of  Eq. (\ref{Z1orb}) are those that we have
already computed in the previous section (apart from an additional
factor $\frac{1}{2}$ coming from the orbifold projection). Here we
need just to evaluate the last two terms. The third one turns out
to be
\begin{eqnarray}
Z_{he}^o & = & \frac{N}{(8 \pi^2 \alpha')^2} \int d^4x \sqrt{
-\mbox{det}(\eta+\hat{F})} \int_{0}^\infty
\frac{d\tau}{\tau} {\rm e}^{-\frac{y^2\tau}{2\pi\alpha'}} \left[ \frac{
4\, \sin \pi \nu_f  \sin \pi \nu_g} {\Theta_{2}^2(0|i\tau)
\Theta_{1}(i\nu_f\tau|i\tau)\Theta_{1} (i\nu_g\tau|i\tau)} \right]
\nonumber \\
{} && \left[ \Theta_{3}^2(0|i\tau)\Theta_{4}(i\nu_f\tau|i\tau)
\Theta_{4}(i\nu_g\tau|i\tau) - \Theta_{4}^2(0|i\tau)
\Theta_{3}(i\nu_f\tau|i\tau) \Theta_{3}(i\nu_g\tau|i\tau) \right]
\nonumber \\
&&+ \frac{iN}{32\pi^2}\int d^4x\, F_{\alpha\beta}^a {\tilde
F}^{a\,\alpha\beta} \int_{0}^\infty
\frac{d\tau}{\tau}
e^{-\frac{y^2 \tau}{2\pi\alpha'}}
\label{zetatet}
\end{eqnarray}
and is equal to the one appearing in the pure orbifold calculation
\cite{LMP}.  To get the previous equation we have used:
\begin{equation}
\label{traccia} {\rm Tr}\langle
ij|e|nm\rangle=\delta_{jj}\delta_{mm}=4N\,\,\,\,\,\,\,\,\,\,\,\,
{\rm Tr}\langle ij|(-1)^{F_S}|nm\rangle=0
\end{equation}
where the indices $i,m = 1,...,N, N+3,...,2N+2 $ enumerate
respectively the
stack of $N$ undressed branes and their images, while the indices
$j,n = N+1, N+2 $ indicate the dressed brane and its image.

Analogously the last term in Eq. (\ref{Z1orb}) can be obtained from
Eq. (\ref{zetatet}) with the substitution $k \rightarrow ik$.
As noticed after Eq. (2.5) of Ref.~\cite{LMP}, in the twisted
sector, only the $NS$ and $NS(-1)^F$ and $R(-1)^F$ sectors
contribute to the interaction . However, the presence of the Type
$0$B projector $\frac{1+(-1)^{F_s}}{2}$ makes the  $NS$ and
$NS(-1)^F$ contributions to vanish because of Eqs. (\ref{cp03})
and (\ref{cp103}). Thus the only twisted sector which gives a non
vanishing contribution to the interaction is the $R(-1)^F$ which is
equal to
\begin{eqnarray}
Z_{h\Omega'I_6}^o= \pm \frac{2i }{32\pi^2}\int d^4x\,
F_{\alpha\beta}^a {\tilde F}^{a\,\alpha\beta} \int_{0}^\infty
\frac{d\tau}{\tau}
e^{-\frac{y^2 \tau}{2\pi\alpha'}} .
\label{teta2}
\end{eqnarray}
The overall factor $2$ again takes in account the two inequivalent
configurations that we have to consider in evaluating the trace,
as discussed after Eq. (\ref{nufg}).
Therefore the coefficient of the kinetic term of the gauge field
in this theory comes only from the second and third term of
Eq. (\ref{Z1orb}) and it turns out to be:
\begin{equation}
\frac{1}{g_{YM}^{2}} = \frac{1}{16 \pi^2} \left[ 2N \mp
\frac{8}{3} \right] \log
  \frac{\mu^2}{\Lambda^2}
\label{runorior}
\end{equation}
consistently with our previous calculation in Eq. (\ref{betaz2}).
Moreover from Eqs. (\ref{zetatet}) and (\ref{teta2}), following the
same procedure as in Ref. \cite{LMP}, one can read also the $\theta$
angle which turns out to be
\begin{eqnarray}
\theta_{YM} = 2\theta(N\pm 2), \label{theta45}
\end{eqnarray}
where  $\theta$ is the phase of the complex cut-off $\Lambda
e^{-i\theta}$.

The gauge theory here obtained shares some common features with
${\cal N}=2$ SYM. As previously noticed, the running coupling
constant and the $\beta$-function of this theory, in the large $N$
limit, reproduce those of ${\cal N}=2$ SYM. Moreover also the
$\theta$ angle in Eq. (\ref{theta45}) in the large $N$ limit
reduces to the one of ${\cal N}=2$ SYM, implying that, in the
planar limit, the two theories are very close to each other. This
connection appears as the natural extension to the case ${\cal
N}=2$ of the Armoni, Shifman and Veneziano planar equivalence for
${\cal N}=1$, in which  the {\em parent} theory is the ${\cal
N}=2$ SYM   and the {\em daughter} theory is the world-volume
theory of $N$ fractional branes of our orbifold.

Finally, it is useful to rewrite the previous expressions in the
closed string channel where,  because of the open/closed string
duality, these amplitudes correspond to the tree level exchange
diagram between a stack of $N$ undressed branes and one brane
dressed with an $SU(N)$ gauge field. In particular, by using
Eqs. (\ref{mod1}) and (\ref{the2}) and the well-known modular
transformation properties  of the $\Theta$-functions, we can
rewrite Eqs. (\ref{zetatet}), (\ref{teta2}) and $Z^o_{\Omega'I_6}$
in the closed string channel. The other terms in the free energy
are irrelevant in the forthcoming discussion because they are
vanishing in the field theory limit. From the annulus we obtain:
\begin{eqnarray}
Z_{he}^{c} & = & \frac{N}{(8 \pi^2 \alpha')^2} \int d^4x \sqrt{
-\mbox{det}(\eta+\hat{F})} \int_{0}^\infty
\frac{dt}{t} {\rm e}^{-\frac{y^2}{2\pi\alpha't}} \left[ \frac{
4\, \sin \pi \nu_f  \sin \pi \nu_g} {\Theta_{4}^2(0|it)
\Theta_{1}(\nu_f|it)\Theta_{1} (\nu_g|it)} \right]
\nonumber \\
{} && \left[ \Theta_{2}^2(0|it)\Theta_{3}(\nu_f |i t)
\Theta_{3}(\nu_g |i t) - \Theta_{3}^2(0|i t)
\Theta_{2}(\nu_f |i t) \Theta_{2}(\nu_g |i t) \right]
\nonumber \\
&&+ \frac{iN}{32\pi^2}\int d^4x\, F_{\alpha\beta}^a {\tilde
F}^{a\,\alpha\beta} \int_{0}^\infty
\frac{d t}{t}
e^{-\frac{y^2 }{2\pi\alpha't}}
\label{zetatetc}
\end{eqnarray}
while from the Moebius strip:
\begin{eqnarray}
Z^{c}_{\Omega'I_6}(F)&=& \mp\frac{1}{(8\pi^2\alpha')^2 } \int
d^4x \sqrt{-{\rm det} (\eta+\hat{F}) } \int_{0}^{\infty}
\frac{dt}{4t^3} {\rm e}^{-\frac{y^2}{2\pi\alpha't}}
\sin\pi\nu_f \sin\pi\nu_g\nonumber\\
&&\frac{ f_2^4(iq) \Theta_2\left(\frac{\nu_f}{2}|it +\frac{1}{2}
\right) \Theta_2 \left(\frac{\nu_g}{2}|it +\frac{1}{2} \right)
}{f_1^4(iq) \Theta_1\left(\frac{\nu_f}{2}|it +\frac{1}{2} \right)
\Theta_1 \left(\frac{\nu_g}{2}|it +\frac{1}{2} \right)}\label{zoi6}\\
Z_{h\Omega'I_6}^c(F)&= &\pm  \frac{2i }{32\pi^2}\int d^4x\,
F_{\alpha\beta}^a {\tilde F}^{a\,\alpha\beta} \int_{0}^\infty
\frac{d t}{t} e^{-\frac{y^2}{2\pi\alpha't}} \,\, . \label{teta2cl}
\end{eqnarray}
By expanding the previous expressions up to quadratic terms in the
external field and isolating only the terms depending on the gauge
field, we have from the annulus:
\begin{eqnarray}
Z_h^c(F)\!\!&\rightarrow&\!\!\left[- \frac{1}{4} \int d^4 x
F_{\alpha \beta}^{a}
F^{a\,\alpha \beta }\right]
 \left\{
- \frac{N}{8 \pi^2}
\int_{0}^{\infty}
\frac{dt}{t} {\rm e}^{-\frac{y^2}{2\pi\alpha't}}
 \right\}  \nonumber\\
&+&  iN \left[ \frac{1}{32\pi^2} \int d^4x
F^a_{\alpha\beta}\tilde F^{a\,\alpha\beta} \right]
\int_{0}^{\infty} \frac{dt}{t}{\rm e}^{-\frac{y^2}{2\pi\alpha't}}
\label{F285}
\end{eqnarray}
and from the Moebius strip:
\begin{eqnarray}
Z_{\Omega'I_6}^c(F)\!\!\!\!&\rightarrow&\!\!\!\!\left[- \frac{1}{4}\! \int\!
d^4 x F_{\alpha \beta}^{a} F^{a\,\alpha \beta }\right]\!\! \left\{\! \pm
\frac{1}{(4\pi)^2}\! \int_{0}^{\infty} \frac{dt}{8t^3}
e^{-\frac{y^2}{2\pi\alpha't} }
\left(
\frac{f_2(iq)}{f_1(iq)}\right)^8 \left[\frac{4}{3} + \frac{1}{
\pi} \partial_t \log f_2^4(iq)\right]\!\! \right\}\nonumber\\
&&\label{teta2cl1}
\end{eqnarray}
Eqs. (\ref{teta2cl}) and (\ref{F285}) are exact at string level
even if they receive contribution only from the massless closed
string states. For $y=0$ both of them are left invariant  under
open/closed string duality  and for this reason  one is  able to
obtain, from the closed channel, the planar contribution to the
$\beta$-function and the complete expression of the  $\theta$-angle.

Eq. (\ref{teta2cl1}), which  in the open channel gives the
subleading behaviour in the large $N$ limit of the
$\beta$-function,  receives, instead, contributions from all the
string states. The massless pole in the open channel is not left
invariant under open/closed string duality and  by performing the
field theory limit on such expression, as explained in the last
section, we obtain a  vanishing result.

In conclusion we can assert that the gauge/gravity correspondence
certainly holds in the planar limit where a Bose-Fermi degeneracy
is recovered and the theory resembles to a supersymmetric theory.
However, some non planar information can  be still obtained from the
closed channel as the example of the $\theta$-angle has showed.

\subsection{Orbifold $C^3/(\mathbb{Z}_2\times \mathbb{Z}_2$) -
A simple string realization
of the Armoni, Shifman and Veneziano model }

Let us consider now the orbifold $C^3/(\mathbb{Z}_2\times
\mathbb{Z}_2)$, which contains four elements $\{1,h_1,h_2,h_3\}$
acting on the three complex coordinates
\begin{eqnarray}
&&z_1=x_4+ix_5 \qquad z_2=x_6+ix_7 \qquad z_3=x_8+ i x_9
\label{z123}
\end{eqnarray}
as follows:
\begin{eqnarray}
R_1= \left[
\begin{array}{ccc}
~~1 & ~0& ~~0
\\ ~~0& ~1 & ~~0 \\ ~~0 & ~0  & ~~1
\end{array}
\right],  && \quad  R_{h_1}=\left[
\begin{array}{ccc}
1 & 0& 0
\\ 0& -1 & 0 \\ 0 & 0  & -1
\end{array}
\right], \nonumber\\
 \quad R_{h_2}= \left[
\begin{array}{ccc}
-1 & 0& 0
\\ 0& 1 & 0 \\ 0 & 0  & -1
\end{array}
\right], &&\quad   R_{h_3}= \left[
\begin{array}{ccc}
-1 & 0& 0
\\ 0& -1 & 0 \\ 0 & 0  & 1
\end{array}
\right] \ .
\end{eqnarray}
and on the spinors as:
\begin{eqnarray}
&&h_1|s_1\, s_2\,s_3\, s_4\rangle=
e^{i \pi (s_3+s_4)}|s_1\, s_2\,s_3\, s_4\rangle \nonumber\\
&&h_2|s_1\, s_2\,s_3\, s_4\rangle=
e^{i \pi (s_2+s_4)}|s_1\, s_2\,s_3\, s_4\rangle  \nonumber\\
&&h_3|s_1\, s_2\,s_4\, s_4\rangle= -e^{i \pi (s_2+s_3)}|s_1\,
s_2\,s_3\, s_4\rangle \label{orbf22}
\end{eqnarray}
where the sign in front of the last equation is required by the group
properties.

The states left invariant are one gauge vector and one Dirac
fermion in the two-index symmetric (or anti-symmetric)
representation of the gauge group. Also in this case the theory
has a common bosonic sector with a supersymmetric model, that is
${\cal N}=1$ SYM. It is simple to check that the $\beta$-function
for this theory is, at one-loop
\begin{eqnarray}
\beta(g_{YM})= \frac{g_{YM}^{3}}{(4 \pi)^2} \left[- \frac{11}{3}N
+ \frac{4}{3} \frac{N \pm 2}{2} \right]=
\frac{g^3_{YM}N}{(4\pi)^2} \left[ -3\pm \frac{4}{3N}\right]
\label{betaz2xz2}
\end{eqnarray}
which differs from the one of ${\cal N}=1$ SYM because of the subleading
term in $1/N$.

Notice that for $N=3$ the two-index antisymmetric representation
is equal to the fundamental one.
Therefore, with the antisymmetric choice, {\it the world-volume
gauge theory living on a stack of $N$ fractional branes in the
orbifold $C^3/(\mathbb{Z}_2\times \mathbb{Z}_2)$ of the
orientifold Type $0B$/ $\Omega' I_6 (-1)^{F_L}$, for $N=3$ is
nothing but one flavour QCD}. This is an alternative and simpler
stringy realization  of the Armoni-Shifman-Veneziano model. In \cite{ASV} 
and reference therein, the same gauge theory is
realized, in the framework of Type $0A$ theory, by considering a
stack of $N$ D$4$ branes on top of an orientifold $O4$ plane,
suspended between orthogonal NS $5$ branes. It would be
interesting to exploit the relation between the two models, which
should be connected through a simple T-duality.

Besides the stringy realization, the gauge theory we end with is
related by planar equivalence to ${\cal N}=1$ SYM. In the language
of Armoni, Shifmann and Veneziano, the symmetric $(+)$ and
antisymmetric $(-)$ choices correspond to the $S$ and $A$
orientifold theories of ${\cal N}=1$ SYM. This opens the way to a
very interesting extension of many predictions of supersymmetric
parent theory to the non supersymmetric daughter theory, which
holds in the large $N$  limit  \cite{ASV}.

 As discussed in Refs.~\cite{NAPOLI}~\cite{FERRO}, the
orbifold $C^3/(\mathbb{Z}_2\times \mathbb{Z}_2)$ can be seen as
obtained by three copies of the orbifold $\mathbb{ C}^2/\mathbb{
Z}_2$ where the $i$-th $\mathbb{Z}_2$ contains the elements $(1,
h_i)$ $(i=1, \dots 3)$.

In particular we consider the interaction between a stack of $N_I$
($I=1, \dots, 4$) branes of type $I$ and a D3-fractional brane of
type  $I=1$ dressed with an $SU(N)$ gauge field. In this case the
interaction turns out to be
given by the sum of eight terms:
\[
Z =  \int_{0}^{\infty} \frac{d \tau}{\tau} Tr_{{\rm
 NS-R}} \left[ \left(\frac{1 +h_1+h_2+h_3}{4}\right)
\left(\frac{e+\Omega'I_6}{2}
\right)\left( \frac{1+  (-1)^{F_s}}{2}\right) \right.
\]
\begin{equation}
{\times} \left.  (-1)^{G_{bc}}\left( \frac{(-1)^{G_{\beta\gamma}}
+ (-1)^F }{2} \right) { e}^{- 2 \pi \tau L_0} \right]
 \equiv  Z_{e}^o + Z_{\Omega'I_6}^o + \sum_{i=1}^{3}
\left[  Z_{h_ie}^o + Z_{h_i\Omega'I_6}^o\right]
\label{Z2orb}
\end{equation}
Here the first two terms are the same as the ones of the previous
orbifolds except for a further factor 1/2 due to the orbifold
projection. The terms $ Z_{h_ie}^o$ turn out to be
\begin{eqnarray}
Z_{h_i}^o & = & \frac{f_i(N)}{2\,(8\pi^2\alpha')^2} \int d^4x
\sqrt{ -\mbox{det}(\eta+\hat{F})} \int_{0}^\infty \frac{d
\tau}{\tau}{\rm e}^{-\frac{y^2\tau}{2\pi\alpha'}}
\frac{4\,\sin\pi\nu_f \sin\pi\nu_g}{ \Theta_{2}^2(0|i \tau)
\Theta_{1}(i \nu_f \tau| i \tau) \Theta_{1}(i \nu_g \tau|i \tau)}
\nonumber\\
&&\left\{\Theta_{3}^2(0|i \tau) \Theta_{4}(i \nu_f \tau|i
\tau)\Theta_{4}(i \nu_g \tau |i \tau) -\Theta_{4}^2(0|i \tau)
\Theta_{3}(i \nu_f \tau|i \tau)
\Theta_{3}(i \nu_g \tau |i \tau)\right\}\nonumber\\
&&+\frac{i\, f_i(N)}{64\pi^2} \int d^4x F^a_{\alpha\beta}\tilde
F^{a \, \alpha\beta}\int_{0}^{\infty} \frac{d\tau}{\tau}{\rm e}^{-\frac{y^2\tau}{2\pi\alpha'}}
\label{ztet1}
\end{eqnarray}
where the  functions $f_i(N)$ depend on the number of 
different kinds of fractional branes $N_I$ and are given by \cite{
NAPOLI}\cite{FERRO}:
\begin{eqnarray}
&& f_1(N_I)= N_1 + N_2 - N_3 - N_4\nonumber\\  && f_2(N_I)= N_1 -
N_2 + N_3 - N_4 \nonumber\\  && f_3(N_I)= N_1 - N_2 - N_3 + N_4 .
\label{coupling}
\end{eqnarray}
As in the previous orbifold case, all the bosonic terms  of
$Z_{h_i\Omega'I_6}$ vanish because of the contribution to the
trace of the  projector $\frac{1+(-1)^{F_s}}{2}$, while the
$R(-1)^F$ sector gives
\begin{eqnarray}
Z_{h_i\Omega'I_6}^o= \pm \frac{2i}{64\pi^2} \int d^4x
F^a_{\alpha\beta}\tilde F^{a \, \alpha\beta}\int
\frac{d\tau}{\tau}{\rm e}^{-\frac{y^2\tau}{2\pi\alpha'}}
\label{teta3}
\end{eqnarray}
By extracting the coefficient of the gauge kinetic term from the
field theory limit of the interaction in Eq. (\ref{Z2orb}) and
specializing to the case $N_1=N,\,N_2=N_3=N_4=0$ we get:
\begin{equation}
\frac{1}{g_{YM}^{2}} = \frac{1}{16 \pi^2} \left[3 N \mp
\frac{4}{3} \right] \log  \frac{\mu^2}{\Lambda^2} ,
\label{runorior1}
\end{equation}
while the theta angle turns out to be
\begin{eqnarray}
\theta_{\rm YM}=(N\pm 2)\theta \label{theta451} .
\end{eqnarray}
We can repeat the same analysis in the closed string channel by
transforming under open/closed string duality Eq. (\ref{Z2orb})
and performing all the steps explained in the
latter section. However, being Eqs. (\ref{ztet1}), (\ref{teta3}) and
$Z_{\Omega'I_6}^o$ coincident, apart from an overall factor, with
Eqs. (\ref{zetatet}) and (\ref{teta2}) we get the same
conclusions as we did in the last subsection.
In the closed string channel one is able to capture,
in the field theory limit, only the planar contribution to the
$\beta$-funtion and the complete expression for the $\theta$-angle.
Gauge/gravity correspondence in these non-supersymmetric models holds only 
in the
large $N$ limit even if some non planar results are still present in
the closed channel.

{\bf{Acknowledgments}}
We thank M Bill\'o, M. Frau, A. Lerda and I. Pesando for
useful discussions and, in particular, one of us (R. M.) warmly thanks 
Universit\`a di Torino for their hospitality. 
A.L., R.M. and F.P. thank Nordita for their kind hospitality in different 
stages of this work.

\appendix

\section{Some useful properties of $\Theta$ functions}
\label{app0}

In this Appendix we give a detailed derivation of Eqs. (\ref{gym03}) and
(\ref{gymc03}).

The $\Theta$-functions which are the solutions of the
heat equation:
\begin{eqnarray}
\frac{\partial}{\partial\tau}
\Theta\left(\nu|i\tau\right)=\frac{1}{4\pi} \partial_{\nu}^2
\Theta(\nu|i\tau) \label{iden}
\end{eqnarray}
are given by:
\begin{eqnarray} &&\Theta_1(\nu
|it)\equiv\Theta_{11}(\nu,|it) =-2 q^{\frac{1}{4}}\sin\pi\nu
\prod_{n=1}^\infty \left[(1-q^{2n}) (1-e^{2i\pi\nu}q^{2n})
(1-e^{-2i\pi\nu}q^{2n}) \right]
\nonumber\\
&&\Theta_2(\nu |it)\equiv\Theta_{10}(\nu,|it) =2
q^{\frac{1}{4}}\cos\pi\nu \prod_{n=1}^\infty \left[(1-q^{2n})
(1+e^{2i\pi\nu}q^{2n})(1+e^{-2i\pi\nu}q^{2n}) \right]
\nonumber\\
&&\Theta_3(\nu,|it)\equiv\Theta_{00}(\nu,|it) =\prod_{n=1}^\infty
\left[(1-q^{2n})(1+e^{2i\pi\nu}{(2n-1)}) (1+e^{-2i\pi\nu}q^{2n-1})
\right]
\nonumber\\
&&\Theta_4(\nu,|it)\equiv\Theta_{01}(\nu,|it) =\prod_{n=1}^\infty
\left[(1-q^{2n})(1-e^{2i\pi\nu}q^{2n-1}) (1-e^{-2i\pi\nu}q^{2n-1})
\right]
\end{eqnarray}
with $q=e^{-\pi t}$. It is also useful to give an alternative
representation of the $\Theta$-functions:
\begin{eqnarray}
\Theta \left[
\begin{array}{ll}
a \\
b
\end{array}\right]
\left(\nu |t\right)
=\sum_{n= -\infty}^{\infty} {e}^{ 2 \pi i \left[ \frac{1}{2} (n +
    \frac{a}{2})^2 t + ( n + \frac{a}{2} )( \nu + \frac{b}{2} \right)]}
\label{theab98}
\end{eqnarray}
where $a,b$ are rational numbers. It is easy to show that
\begin{equation}
\Theta \left[
\begin{array}{ll}
1 \\
1
\end{array}\right]
\left(\nu |t\right) = - i \sum_{n= - \infty}^{\infty} (-1)^n {e}^{i \pi
  t (n-\frac{1}{2})^2} {e}^{i \pi \nu (2n-1)} \equiv
\Theta_1 \left(\nu |t\right)
\label{the198}
\end{equation}
and
\begin{equation}
\Theta \left[
\begin{array}{ll}
1 \\
0
\end{array}\right]
\left(\nu |t\right) =  \sum_{n= - \infty}^{\infty}  {e}^{i \pi
  t (n-\frac{1}{2})^2} {e}^{i \pi \nu (2n-1)} = \Theta_2 \left(\nu |t\right)
\equiv - \Theta_1 \left( \nu + \frac{1}{2} |t\right) .
\label{the298}
\end{equation}
From the definition in Eq. (\ref{theab98}) it is easy to derive the
following identity:
\begin{eqnarray}
\Theta \left[
\begin{array}{ll}
a \\
b
\end{array}\right]
\left(\nu+\frac{\epsilon_1}{2}t+\frac{\epsilon_2}{2}|t\right)
=e^{-\frac{i\pi t\epsilon_1^2}{4}}
e^{-\frac{i\pi \epsilon_1}{2}(2\nu+b)} e^{-\frac{i\pi
\epsilon_1\epsilon_2}{2}}\Theta \left[
\begin{array}{ll}
a +\epsilon_1 \\
b +\epsilon_2
\end{array}\right](\nu|t) .
\label{theide98}
\end{eqnarray}

In order to get the Euler-Heisenberg action in Eq. (\ref{263}) we
need to use the following expressions which hold for
$\tau\rightarrow\infty$ and $\alpha'\rightarrow 0$ :
\begin{equation}
\Theta_1 ( i \nu \tau| i \tau + \frac{1}{2} ) \rightarrow - 2 i
(ik)^{1/4}  \sinh \pi \nu \tau~~,~~ \Theta_2 ( i \nu \tau| i \tau
+ \frac{1}{2} ) \rightarrow  2 (ik)^{1/4}  \cosh \pi \nu \tau
\label{the12}
\end{equation}
\begin{equation}
\nu_f \rightarrow - 2 \alpha' i {\hat{f}}~~,~~ \nu_g \rightarrow -
2 \alpha'  {\hat{g}}
\label{nuf}
\end{equation}
and
\begin{equation}
f_1 (ik) \rightarrow (ik)^{1/12}~~~,~~~f_2 (ik) \rightarrow
\sqrt{2} (ik)^{1/12}
\label{flim89}
\end{equation}
which, together with
\begin{equation}
\sqrt{-{\rm det} (\eta+\hat{F}) } \sin \pi \nu_f \sin \pi \nu_g =
i (2 \pi \alpha')^2 {\hat{f}}{\hat{g}}~~~,
\label{equa23}
\end{equation}
leads to Eq. (\ref{263}).

In order to derive Eq. (\ref{gym03}) we need the following expansions
of the $\Theta$-functions up to the quadratic order in the gauge fields:
\begin{eqnarray}
\Theta_n\left[i\nu_f\tau|i\tau+1/2\right]& \simeq&
\Theta_n\left[0|i\tau+1/2\right] +2
\frac{\tau^2}{\pi}\partial_\tau
\Theta_n\left[0|i\tau+1/2\right]f^2\nonumber\\
&=& f_1(ik)f_n^2(ik) \left[1+ 2\frac{\tau^2}{\pi}
\partial_\tau \log [f_1(ik)f_n^2(ik)]\, f^2\right]
\label{thex}
\end{eqnarray}
for  $n=2,3,4$ and
\begin{eqnarray}
\frac{\sin\pi\nu_f}{\Theta_1\left[i\nu_f\tau|i\tau+1/2\right]}\simeq
\frac{i}{2\tau f_1^3(ik)}\left[ 1+\left(\frac{1}{6}+ \tau^2
k\frac{\partial}{\partial k} \log f_1^2(ik)\right)f^2\dots\right]
\label{th1ex}
\end{eqnarray}
for $\Theta_1$. Inserting them in Eq. (\ref{ZF03}) we get $(k=
{e}^{-\pi \tau})$:
\begin{eqnarray}
Z^F&\simeq& \pm 4 \frac{1}{(8\pi^2\alpha')^2} \int d^4 x
\int_{0}^{\infty} \frac{d\tau}{\tau}
e^{-\frac{y^2\tau}{2\pi\alpha'} } \left[1- \frac{1}{2} (f^2-g^2)
\right] \left[\frac{i}{ 2 \tau f_1^3 (ik)} \right]^2
\left[  \frac{f_2^8(ik)}{f_1^2(ik)}\right]\nonumber\\
&{\times}&\!\!\left[1 + \left(\frac{1}{6} +\tau^2k
    \frac{\partial}{\partial k}
\log  f_1^2(ik) \right) (f^2-g^2)\right] \left[ 1+2(f^2-g^2)\frac{
\tau^2}{\pi} \partial_\tau
\log  \left( f_1(ik) f_2^2(ik) \right) \right]\nonumber\\
&=&\mp \frac{1}{(8\pi^2\alpha')^2} \int d^4 x \int_{0}^{\infty}
\frac{d\tau}{\tau} e^{-\frac{y^2\tau}{2\pi\alpha'} }
\left(  \frac{f_2(ik)}{f_1(ik)}\right)^8\nonumber\\
&{\times}& \left[\frac{1}{\tau^2} + \left(- \frac{1}{3\tau^2} +
\frac{2}{\pi} \partial_\tau\log f_2^2(ik)\right) (f^2-g^2)\right]
\end{eqnarray}
From it we can easily obtain Eq. (\ref{gym03}) if we take into account
that $f (g) = 2 \pi \alpha' {\hat{f}} ({\hat{g}})$.

In order to write the interaction in Eq. (\ref{ZF03}) in the closed
channel we need to perform the modular transformation $\tau=1/4t$ that
gives
\begin{eqnarray}
Z^{c}(F)_{\Omega'I_6}&=&\mp \frac{1}{(8\pi^2\alpha')^2 } \int d^4x
\sqrt{-{\rm det} (\eta+\hat{F}) } \int_{0}^{\infty} \frac{dt}{t^3}
\sin\pi\nu_f \sin\pi\nu_g\nonumber\\
&&\frac{ f_2^4(ie^{-\pi t})
\Theta_2\left(i\frac{\nu_f}{4t}|\frac{i}{4t} +\frac{1}{2} \right)
\Theta_2 \left(i\frac{\nu_g}{4t}|\frac{i}{4t}+\frac{1}{2} \right)
}{f_1^4(ie^{-\pi t})
\Theta_1\left(i\frac{\nu_f}{4t}|\frac{i}{4t}+\frac{1}{2} \right)
\Theta_1 \left(i\frac{\nu_g}{4t}|\frac{i}{4t}+\frac{1}{2}\right)} .
\label{ZF04}
\end{eqnarray}
In the following we will write the formulas for the $\Theta$-functions
that are needed  to get the previous equation and to show that it is
equal to Eq. (\ref{ZF05}).

Under an arbitrary modular transformation $t\rightarrow
\frac{at+b}{ct+d}$,  $\Theta_1$ transforms as follows
\begin{equation}
\Theta_1\left(\frac{\nu}{ct+d}|\frac{at+b}{ct+d}\right)
=\eta'\,\Theta_{1}({\nu} |t)e^{i\pi c\nu^2 /(ct+d)}
(ct+d)^{\frac{1}{2}} \label{genmod}
\end{equation}
where $\eta'$ is an eight root of unity. It implies
the following transformation of
$\Theta_1$:
\begin{eqnarray}
\Theta_1\left(-\frac{\nu}{2} |\frac{t}{4}-\frac{1}{2}\right)
=\frac{1}{\eta'}\,\Theta_{1}\left(-\frac{\nu}{t}
|\frac{1}{2}-\frac{1}{t}\right)e^{-i\pi \nu^2 /t}
\left(\frac{2}{t}\right)^{\frac{1}{2}}
\label{mod1a}
\end{eqnarray}
that is obtained from Eq. (\ref{mod1a}) by first making the
 substitutions $ \nu \rightarrow - \frac{\nu}{2}$ and $ t \rightarrow
 \frac{t}{4} - \frac{1}{2}$ and then choosing $ a=d=1, c=2$ and
 $b=0$. By performing in the previous equation the substitution $ t
 \rightarrow 4i t$ we get the following equation:
\begin{equation}
\Theta_1\left(-\frac{\nu}{2} |it -\frac{1}{2}\right)
=\frac{1}{\eta'}\,\Theta_{1}\left( \frac{i\nu}{4t}
|\frac{1}{2}+\frac{i}{4t}\right)e^{-\pi \nu^2 /(4t)}
\left(\frac{1}{2it}\right)^{\frac{1}{2}} .
\label{the1a}
\end{equation}
Finally, by using Eq. (\ref{genmod}) with $a=b=d=1$ and $c=0$ we can write:
\begin{eqnarray}
\Theta_1\left(\nu| t +1\right)=\eta' \Theta_1\left(\nu| t \right) .
\label{mod11}
\end{eqnarray}
The latter allows us to write
\begin{eqnarray}
\Theta_1\left(-\frac{\nu}{2} |it -\frac{1}{2}\right)=\frac{1}{\eta'}
\Theta_1\left(-\frac{\nu}{2} |it +\frac{1}{2}\right)
\label{mod33}
\end{eqnarray}
that inserted in Eq. (\ref{the1a}) leads to
\begin{eqnarray}
\Theta_1\left(-\frac{\nu}{2} |it +\frac{1}{2}\right)
=\Theta_{1}\left( \frac{i\nu}{4t}
|\frac{1}{2}+\frac{i}{4t}\right)e^{-\pi \nu^2 /(4t)}
\left(\frac{1}{2it}\right)^{\frac{1}{2}} .
\label{mod1}
\end{eqnarray}

In order to get the analogous
transformation property of $\Theta_2$ we use the following
relation:
\begin{eqnarray}
\Theta_2\left(-\frac{\nu}{2} |\frac{t}{4}-\frac{1}{2}\right)
=-\Theta_{1}\left(\frac{1-\nu}{2}|\frac{t}{4}-\frac{1}{2}\right) .
\label{mod2}
\end{eqnarray}
Then, by applying Eq. (\ref{genmod})  with
$\nu\rightarrow\frac{1-\nu}{2}$,
$t\rightarrow\frac{t}{4}-\frac{1}{2}$ and  $a=d=1,\,c=2,\,b=0$ ,
to the second term of the previous equation, it can be rewritten
as
\begin{eqnarray}
\Theta_1\left(\frac{1-\nu}{2} |\frac{t}{4}-\frac{1}{2}\right)
=\frac{1}{\eta'}\,\left(\frac{2}{t}\right)^{1/2}\,e^{-\frac{i\pi(1-\nu)^2}{
t}}\,
\Theta_{1}\left(\frac{1-\nu}{t}\,|\frac{1}{2}-\frac{1}{t}\right)
\end{eqnarray}
and then substituting it in  Eq. (\ref{mod2}) and  performing the
substitution  $t\rightarrow 1 /t$, one gets:
\begin{eqnarray}
\Theta_2\left(-\frac{\nu}{2} |\frac{1}{4t}-\frac{1}{2}\right)
=-\frac{1}{\eta'}\,(2t)^{1/2}\,e^{-i\pi(1-\nu)^2 t}\,
\Theta_{1}\left((1-\nu) t\,|\frac{1}{2}-t\right) . \label{mod2b}
\end{eqnarray}
Let us consider the $\Theta_1$ appearing in the second term of the
previous equation. By defining in it $t'\equiv\frac{1}{2}-t$ and
then $\nu'\equiv -\nu\left(\frac{1}{2}-t'\right)$ we can rewrite
it as
\begin{eqnarray}
\Theta_{1}\left((1-\nu)t\,|\frac{1}{2}-t\right)=
\Theta_{1}\left(\nu'-t'+\frac{1}{2}\,|t'\right). \label{mod2cb}
\end{eqnarray}
Therefore, by using Eq. (\ref{theide98})
with $a=b=1$, $\epsilon_2=1, \epsilon_1=-2$, we can
write Eq. (\ref{mod2cb}) as follows:
\begin{eqnarray}
\Theta_{1}\left((1-\nu) t\,|\frac{1}{2}-t\right)=i
e^{i\pi(1-2\nu) t} \Theta_{2}\left(-\nu t\,|\frac{1}{2}-t\right)
\label{mod2bb}
\end{eqnarray}
where we have restored the variables $\nu$ and $t$ and used the following identity:
\begin{eqnarray}
\Theta \left[
\begin{array}{ll}
-1 \\
\,\,\,\,2
\end{array}\right](\nu|t)=-\Theta \left[
\begin{array}{ll}
1 \\
0
\end{array}\right](\nu|t)=-\Theta_2(\nu|t)
\end{eqnarray}
that can be easily derived starting from the general expression of the theta-function given in Eq.
(\ref{theab98}). Then by
inserting  Eq. (\ref{mod2bb}) in Eq. (\ref{mod2b}) we get
\begin{eqnarray}
\Theta_2\left(-\frac{\nu}{2} |\frac{1}{4t}-\frac{1}{2}\right)
=-\frac{i}{\eta'}\,\Theta_{2}\left(-\nu t
|\frac{1}{2}-t\right)e^{-i\pi \nu^2 t} (2t)^{\frac{1}{2}} .
\label{mod4}
\end{eqnarray}
Furthermore, by performing the substitution $ t \rightarrow -
\frac{i}{4t}$, Eq. (\ref{mod4})  becomes
\begin{eqnarray}
\Theta_2\left(-\frac{\nu}{2} | {it} -\frac{1}{2}\right)
=-\frac{i}{\eta'}\,\Theta_{2}\left(\frac{i \nu}{4 t} |\frac{1}{2} +
\frac{i}{4t} \right)e^{- \pi \nu^2 /(4t)} (2it)^{-\frac{1}{2}} .
\label{the2a}
\end{eqnarray}
Finally, by rewriting  $\Theta_2$-function in terms of $\Theta_1$ by
means of Eq. (\ref{the298}) and using Eq. (\ref{mod11}),
we can write:
\begin{eqnarray}
\Theta_2\left(-\frac{\nu}{2} | {it} -\frac{1}{2}\right)=
\frac{1}{\eta'}
\Theta_2\left(-\frac{\nu}{2} | {it} +\frac{1}{2}\right) .
\label{pippo}
\end{eqnarray}
The last identity allows us to write:
\begin{eqnarray}
\Theta_2\left(-\frac{\nu}{2} | {it} +\frac{1}{2}\right)
=-i\Theta_{2}\left(\frac{i \nu}{4 t} |\frac{1}{2} +
\frac{i}{4t} \right)e^{- \pi \nu^2 /(4t)} (2it)^{-\frac{1}{2}} .
\label{the2}
\end{eqnarray}
The modular transformations of the $f$-functions with complex
argument are:
\begin{eqnarray}
f_1(ie^{-\pi s})=(2s)^{-1/2}f_1(ie^{-\frac{\pi}{4 s}})&&
f_2(ie^{-\pi
  s})=f_2(ie^{-\frac{\pi}{4 s}})\label{f1f2}\\
f_3(ie^{-\pi s})=\,\,\,\,\,\,\,e^{i\pi/8}f_4(ie^{-\frac{\pi}{4
s}})&& f_4(ie^{-\pi s})=e^{-i\pi/8} f_3(ie^{-\frac{\pi}{4 s}})
\label{f3f4}
\end{eqnarray}
Eq. (\ref{ZF04}) is obtained from Eq. (\ref{ZF03}) by using
Eqs. (\ref{f1f2}) and by changing variable from $\tau$ to $t =
\frac{1}{4 \tau}$. Finally by using Eqs. (\ref{mod1}) and (\ref{the2})
one gets Eq. (\ref{ZF05}) from Eq. (\ref{ZF04}).

In order to obtain Eq. (\ref{gymc03}) we have used the following
expansions in the external field
\begin{eqnarray}
\Theta_n\left(\frac{\nu_f}{2} |it+\frac{1}{2}\right)& \simeq&
\Theta_n\left[0|it+\frac{1}{2}\right] -\frac{2}{\pi}\partial_t
\Theta_n\left(0|it+\frac{1}{2}\right)\frac{f^2}{4}\nonumber\\
&=& f_1(iq)f_n^2(iq) \left[1+ 2q\partial_q \log
[f_1(iq)f_n^2(iq)]\, \frac{f^2}{4}\right]
\label{thexcl}
\end{eqnarray}
for $n=2,3,4$, and
\begin{eqnarray}
\frac{\sin\pi\nu_f}{\Theta_1\left(\frac{\nu_f}{2}|it+\frac{1}{2}\right)}\simeq
-\frac{1}{f_1^3(iq)}\left\{
1+f^2\left[\frac{1}{8}-q\partial_q\frac{1}{2}
\log\prod_n\left(1-(iq)^{2n}\right)\right]\right\} \label{th1excl}
\end{eqnarray}
for $\Theta_1$.

\end{document}